\documentclass[
final,
leqno,
]{siamltex1213}

\usepackage{pstricks}
\usepackage{amsfonts}
\usepackage{amsmath}
\usepackage{mathtools}
\usepackage{graphicx}
\usepackage{wrapfig}
\usepackage{alg}
\usepackage{amssymb}
\usepackage{color}
\usepackage{enumerate}
\usepackage[separate-uncertainty=true]{siunitx}
\usepackage{listings}
\usepackage{tikz}
\usepackage{pgfplots}
\usepackage{pgfplotstable}
\usepackage{array}
\usepackage{booktabs}
\usepackage{placeins}
\usepackage{subfigure}
\usepackage[toc,page]{appendix}
\usepackage{siunitx}
\usepackage{multirow}
\usepackage{caption}
\usepackage{longtable}

\pgfkeys{/pgfplots/styleA/.style={smooth,draw=black,mark=*,mark options={fill=green!60!black,draw=black}}}
\pgfkeys{/pgfplots/styleA2/.style={smooth,draw=black,mark=*,mark options={fill=blue!60!black,draw=black}}}
\pgfkeys{/pgfplots/styleB/.style={smooth,densely dashed,draw=black,mark=*,mark options={solid,fill=yellow!70!black,draw=black}}}
\pgfkeys{/pgfplots/speedup style/.style={draw=green!40!black}}
\pgfkeys{/pgfplots/speedup2 style/.style={draw=blue!40!black}}

\newcommand{\SpikeHyb}{\tt SaP::GPU}
\newcommand{\SaP}{\tt SaP}
\newcommand{\SaPC}{\tt SaP::GPU-C}
\newcommand{\SaPD}{\tt SaP::GPU-D}
\newcommand{\DB}{\tt DB}
\newcommand{\DBS}[1]{\tt DB-S{#1}}
\newcommand{\CM}{\tt CM}
\newcommand{\RCM}{\tt RCM}
\newcommand{\CMS}[1]{\tt CM-S{#1}}

\newcommand{\BFS}{\tt BFS}
\newcommand{\TSR}{3$^{rd}$ SR}

\newcommand{\BCG}{\tt BiCGStab}

\newcommand{\BCGtwo}{\tt BiCGStab(2)}
\newcommand{\MCs}{\tt MC60}
\newcommand{\MCsf}{\tt MC64}
\newcommand{\Pardiso}{\tt PARDISO}
\newcommand{\SuperLU}{\tt SuperLU}
\newcommand{\MUMPS}{\tt MUMPS}
\newcommand{\cuSOLVER}{\tt cuSOLVER}
\newcommand{\Paralution}{\tt Paralution}
\newcommand{\MKL}{\tt MKL}

\newcommand{\SciNum}[1]{\num[round-precision=4,round-mode=figures,scientific-notation=true,output-exponent-marker=\ensuremath{\mathrm{E}}]{#1}}

\newcommand{\R}{\mathbb{R}}

\newcommand{\bA}{{\bf A}}
\newcommand{\bB}{{\bf B}}
\newcommand{\bC}{{\bf C}}
\newcommand{\bD}{{\bf D}}

\newcommand{\bL}{{\bf L}}
\newcommand{\bM}{{\bf M}}
\newcommand{\bN}{{\bf N}}

\newcommand{\bQ}{{\bf Q}}
\newcommand{\bR}{{\bf R}}
\newcommand{\bS}{{\bf S}}
\newcommand{\bU}{{\bf U}}
\newcommand{\bV}{{\bf V}}
\newcommand{\bW}{{\bf W}}
\newcommand{\zero}{{\bf 0}}

\newcommand{\bb}{{\bf b}}

\newcommand{\bg}{{\bf g}}

\newcommand{\bx}{{\bf x}}

\newcommand{\T}[2]{{#1}_{#2}^{(t)}}
\newcommand{\M}[2]{{#1}_{#2}^\prime}
\newcommand{\B}[2]{{#1}_{#2}^{(b)}}

\title{
ANALYSIS OF A SPLITTING APPROACH FOR THE PARALLEL SOLUTION OF LINEAR SYSTEMS ON GPU CARDS\thanks{This work was supported by the National Science Foundation grant SI2-SSE--1147337}
}

\author{
Ang Li\footnotemark[2]
\and 
Radu Serban\footnotemark[3] 
\and 
Dan Negrut\footnotemark[3]\footnotemark[2]
}

\begin{document}
\maketitle    

\renewcommand{\thefootnote}{\fnsymbol{footnote}}

\footnotetext[2]{Electrical and Computer Engineering, University of Wisconsin--Madison, Madison, WI 53706}
\footnotetext[3]{Mechanical Engineering, University of Wisconsin--Madison, Madison, WI 53706}

\renewcommand{\thefootnote}{\arabic{footnote}}

\slugger{sisc}{xxxx}{xx}{x}{x--x}

\begin{abstract}
We discuss an approach for solving sparse or dense banded linear systems $\bA \bx = \bb$ on a Graphics Processing Unit (GPU) card. The matrix $\bA \in {\mathbb{R}}^{N \times N}$ is possibly nonsymmetric and moderately large; i.e., $\SI{10000}{} \leq N \leq \SI{500000}{}$. The \emph{split and parallelize} ({\SaP}) approach seeks to partition the matrix $\bA$ into diagonal sub-blocks $\bA_i$, $i=1,\ldots,P$, which are independently factored in parallel. The solution may choose to consider or to ignore the matrices that couple the diagonal sub-blocks $\bA_i$. This approach, along with the Krylov subspace-based iterative method that it preconditions, are implemented in a solver called {\SpikeHyb}, which is compared in terms of efficiency with three commonly used sparse direct solvers: {\Pardiso}, {\SuperLU}, and {\MUMPS}. {\SpikeHyb}, which runs entirely on the GPU except several stages involved in preliminary row-column permutations, is robust and compares well in terms of efficiency with the aforementioned direct solvers. In a comparison against Intel's {\MKL}, {\SpikeHyb} also fares well when used to solve dense banded systems that are close to being diagonally dominant. 
{\SpikeHyb} is publicly available and distributed as open source under a permissive BSD3 license.
\end{abstract}



\pagestyle{myheadings}
\thispagestyle{plain}
\markboth{A. Li, R. Serban, and D. Negrut}{GPU-BASED PARALLEL LINEAR SOLVER}

%=============================================================================

%Where are we going to submit this paper? SISC is the current target, under the "Software and High-Performance Computing" section.

%==============================================================================

\section{Introduction}
\label{s:intro}
Previously used in niche applications and by a small group of enthusiasts, general purpose computing on graphics processing unit (GPU) cards has gained widespread popularity after the release in 2007 of the CUDA programming environment \cite{cudaProgGuide}. Owing also to the release of the OpenCL specification \cite{openCL} in 2008, GPU computing has been rapidly adopted by numerous groups with computing needs originating in a broad spectrum of application areas. 
In several of these areas though, when compared to the library ecosystem enabling sequential and/or parallel computing on x86 chips, GPU computing library support continues to be spotty. This observation motivated an effort whose outcomes are reported in this paper, which is concerned with solving sparse linear systems of equations on the GPU. 

Developing an approach and implementing parallel code for solving sparse linear systems is not trivial. This, and the relative novelty of GPU computing explain the scarcity of solutions for solving $\bA \bx = \bb$ on the GPU, when $\bA \in {\mathbb{R}}^{N \times N}$ is possibly nonsymmetric, sparse, and moderately large; i.e., $\SI{10000}{} \leq N \leq \SI{500000}{}$. An inventory of software solutions as of 2015 produced a short list of codes that solved $\bA \bx = \bb$ on the GPU: {\cuSOLVER}~\cite{cuSOLVER}, {\Paralution}~\cite{paralution}, and {\SuperLU}~\cite{demmel2011superlu}, the latter focused on distributed memory architectures and leveraging GPU computing at the node level only. 
Several CPU multi-core approaches exist and are well established, see for instance \cite{HSL,schenk2004solving,mumps,demmel2011superlu}. For a domain-specific application implemented on the GPU that calls for solving ${\bf A} {\bf x} = {\bf b}$, one alternative would be to fall back on one of these CPU-based solutions. This strategy usually impacts the overall performance of the algorithm due to the back-and-forth data movement across the PCI host--device interconnect, which in practice supports bandwidths of the order of 10 GB/s. Herein, the focus is not on this strategy. Instead, we are interested in carrying out the LU factorization on the GPU when the possibly nonsymmetric matrix ${\bf A}$ is sparse or dense banded with narrow bandwidth.

There are pros and cons to having a linear solver on the GPU. On the upside, since a parallel implementation of a LU factorization is memory bound, particularly for sparse systems, the GPU is attractive owing to its high bandwidths and relatively low latencies. At main-memory bandwidths of roughly 300 GB/s, the GPU is four to five times faster than a modern multicore CPU. On the downside, the irregular memory access patterns associated with sparse matrix factorization ablate this GPU-over-CPU advantage, which is further eroded by the intense logic and integer arithmetic requirements associated with existing algorithms. The approach discussed herein alleviates these two pitfalls by embracing a splitting strategy described for CPU-centric multicore and/or multi-node computing in \cite{PoSa2006}. Two successive row--column permutations attempt to increase the diagonal dominance of the matrix and reduce its bandwidth, respectively. Ideally, the reordered matrix would be ($i$) diagonal dominant, and ($ii$) dense banded. If ($i$) is accomplished, no LU factorization row/column pivoting is necessary, thus avoiding tasks at which the GPU does not shine: logic and arithmetic operations. Additionally, if ($ii$) holds, coalesced memory access patterns associated with dense matrix operations can capitalize on the GPU's high bandwidth. 

The overall solution strategy adopted herein solves ${\bf A} {\bf x} = {\bf b}$ using a Krylov-subspace method and employs LU preconditioning with work-splitting and drop-off. Specifically, each outer Krylov-subspace iteration takes at least one preconditioner solve step that involves solving ${\hat {\bf A}} {\bf y} = {\hat{\bf b}}$ on the GPU, where ${\hat {\bf A}} \in \R^{N \times N}$ is a {\em dense} banded matrix obtained from ${\bf A}$ after a sequence of possibly two reordering stages that can include element drop-off. 
Regardless of whether ${\bf A}$ is sparse or not, the salient attribute of the approach is the casting of the preconditioning step as a {\em dense} linear algebra problem.
Thus, a reordering process is employed to obtain a narrow--band, dense ${\hat {\bf A}}$, which is subsequently LU--factored. For the reordering, a strategy that combines two stages, namely diagonal dominance boosting and bandwidth reduction, has yielded well balanced coefficient matrices that can be factored fast on the GPU leveraging a single instruction multiple data (SIMD)--friendly underlying data structure. 
The LU factorization relies on a splitting of the matrix ${\hat {\bf A}}$ in several diagonal blocks that are factored independently and a correction process to account for the inter-diagonal block coupling.
The implementation takes advantage of the GPU's deep memory hierarchy, its multi-SM layout, and its predilection for SIMD computation. 

This paper is organized as follows. Section \ref{s:description} summarizes the solution algorithm. The discussion covers first the work-splitting-based LU factorization of dense banded matrices. Subsequently, the ${\bf A} {\bf x} = {\bf b}$ sparse case brings into focus strategies for matrix reordering. Section \ref{s:implementation} summarizes aspects related to the GPU implementation of the solution approaches proposed. Results of a series of numerical experiments for both dense banded and sparse linear systems are reported in Section \ref{s:experiments}. Since reordering strategies play a pivotal role in the sparse linear system solution, we present benchmarking results in which we compared the reordering strategies adopted herein to established solutions/implementations. The paper concludes with a series of final remarks and a summary of lessons learned and directions of future work.

%==============================================================================
\section{Description of the methodology}
\label{s:description}

\subsection{The dense banded linear system case}
\label{ss:denseLinSysExp}
\begin{figure}[ht]
\centering {\includegraphics[width=0.85\textwidth]{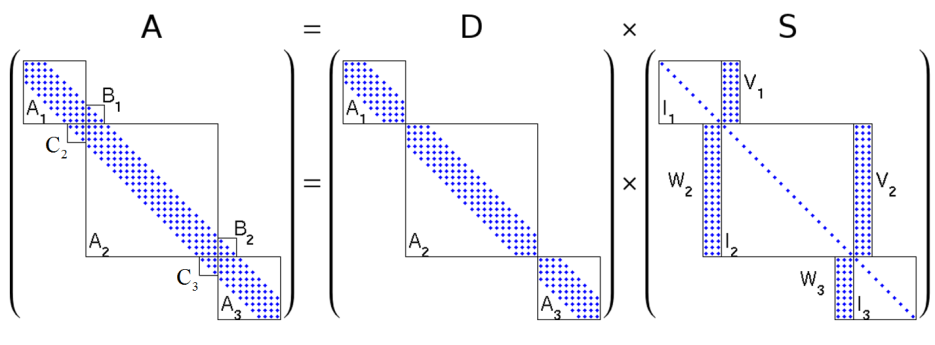}}
\caption{Factorization of the matrix $\bA$ with $P=3$.}
\label{f:matrix_partitioning}
\end{figure}

Assume that the banded dense matrix ${\bA}\in{\mathbb{R}}^{N\times N}$ has half-bandwidth $K\ll N$. 
Following an approach discussed in \cite{SaKu1978, PoSa2006, PoSa2007}, we partition the banded matrix $\bA$ into a block tridiagonal form with $P$ diagonal blocks $\bA_i \in \R^{N_i \times N_i}$, where $\sum_i^P N_i = N$. For each partition $i$, let $\bB_i$, $i=1,\ldots,P-1$ and $\bC_i$, $i=2,\ldots,P$ be the super- and sub-diagonal coupling blocks, respectively -- see Figure \ref{f:matrix_partitioning}. Each coupling block has dimension $K\times K$ for banded matrices with half-bandwidth $K=\max\limits_{i,j,a_{ij} \ne 0}|i - j|$.

As illustrated in Fig.~\ref{f:matrix_partitioning}, the banded matrix $\bA$ is expressed as the product of a block diagonal matrix $\bD$ and a so-called {\em spike matrix} $\bS$ \cite{SaKu1978}. The latter is made up of identity diagonal blocks of dimension $N_i$, and off-diagonal spike blocks, each having $K$ columns. Specifically,
\begin{equation}
\bA = \bD \bS \, ,
\end{equation}
where $\bD = \mbox{diag}(\bA_1,\ldots,\bA_P)$ and, assuming that $\bA_i$ are non-singular, the so-called left and right spikes $\bW_i$ and $\bV_i$ associated with partition $j$, 
each of dimension $N_i \times K$, are given by
\begin{subequations}\label{eq:LeftRightSpikes}
% Calculating V_1
\begin{alignat}{2}
\label{eq:RightSpike_1}
\bA_1 \bV_1 &= \left[\begin{matrix} \zero \\ \zero \\ \bB_1 \end{matrix}\right]   &  &{}\\
% Calculating V_i and W_i
\label{eq:LeftRightSpikes_i}
\bA_i \left[  \bW_i \mid \bV_i \right] &= \left[\begin{matrix}  
\bC_i & \zero \\ \zero & \zero \\ \zero & \bB_i
\end{matrix}\right] \, ,&   &\quad i= 2, \ldots, P-1\\ 
% Calculating W_P
\label{eq:LeftSpike_P}
\bA_P \bW_P &=  \left[\begin{matrix} \bC_P \\ \zero \\ \zero \end{matrix}\right] . & &{}
\end{alignat}
\end{subequations}

% -----------------
% -----------------

Solving the linear system $\bA \bx = \mathbf{b}$ is thus reduced to solving 
\begin{align}
\bD \bg &= \mathbf{b} \label{eq:diag-sys} \\
\bS \bx &= \bg \label{eq:spike-sys}
\end{align}
Since $\bD$ is block-diagonal, solving for the modified right-hand side $\bg$ from (\ref{eq:diag-sys}) is trivially parallelizable, as the work is split across $P$ processes, each charted to solve $\bA_i \bg_i = \bb_i$, $i=1,\ldots,P$. Note that the same decoupling is manifest in Eq.~(\ref{eq:LeftRightSpikes}), and the work is spread over $P$ processes.

The remaining question is how to solve quickly the linear system in (\ref{eq:spike-sys}). This problem can be reduced to one of smaller size, $\hat\bS \hat\bx = \hat\bg$. To that end, the spikes $\bV_i$ and $\bW_i$, as well as the modified right-hand side $\bg_i$ and the unknown vectors $\bx_i$ in (\ref{eq:spike-sys}) are partitioned into their top $K$ rows, the middle $N_i - 2K$ rows, and the bottom $K$ rows:
\begin{subequations}
\begin{alignat}{2}
\bV_i &= \left[\begin{matrix} \T{\bV}{i} \\ \M{\bV}{i} \\ \B{\bV}{i}\end{matrix}\right], & \quad
\bW_i &= \left[\begin{matrix} \T{\bW}{i} \\ \M{\bW}{i} \\ \B{\bW}{i}\end{matrix}\right],  \\
\bg_i &= \left[\begin{matrix} \T{\bg}{i} \\ \M{\bg}{i} \\ \B{\bg}{i}\end{matrix}\right], & \quad
\bx_i &= \left[\begin{matrix} \T{\bx}{i} \\ \M{\bx}{i} \\ \B{\bx}{i}\end{matrix}\right].
\end{alignat}
\end{subequations}
A block-tridiagonal reduced system is obtained by excluding the middle partitions of the spike matrices as:
\begin{equation}\label{eq:SPIKE_reduced_system}
\left[\begin{matrix}
\bR_1 & \bM_1 &  &  &  \\
 & \ddots &  & & \\
 & \bN_i & \bR_i & \bM_i & \\
 & &  & \ddots & \\
& & & \bN_{P-1} & \bR_{P-1}
\end{matrix}\right]
\left[\begin{matrix}
\hat\bx_1 \\ \vdots \\ \hat\bx_i \\ \vdots \\ \hat\bx_{P-1}
\end{matrix}\right] 
=
\left[\begin{matrix}
\hat\bg_1 \\ \vdots \\ \hat\bg_i \\ \vdots \\ \hat\bg_{P-1}
\end{matrix}\right] ,
\end{equation}
where the linear system above, denoted $\hat\bS \hat\bx = \hat\bg$, is of dimension $2K(P-1) \ll N$,
\begin{subequations}
\begin{alignat}{2}
\bN_i & = \left[\begin{matrix}
\B{\bW}{i} & \zero \\ \zero & \zero
\end{matrix}\right] \, ,&  &\quad i=2,\ldots,P-1 \\
\bR_i &= \left[\begin{matrix}
{\bf I}_M & \B{\bV}{i} \\ \T{\bW}{i+1} & {\bf I}_M
\end{matrix}\right] \, ,&  &\quad i=1,\ldots,P-1 \\
\bM_i &= \left[\begin{matrix}
\zero & \zero \\  \zero & \T{\bV}{k+1}
\end{matrix}\right] \, ,&  &\quad i=1,\ldots,P-2
\end{alignat}
\end{subequations}
and
\begin{equation}
\hat\bx_i = \left[\begin{matrix} \B{\bx}{i} \\ \T{\bx}{i+1} \end{matrix}\right] , \,
\hat\bg_i = \left[\begin{matrix} \B{\bg}{i} \\ \T{\bg}{i+1} \end{matrix}\right] , \quad
i=1,\ldots,P-1 \; .
\end{equation}

Two strategies are proposed in \cite{PoSa2006} to solve (\ref{eq:SPIKE_reduced_system}): ({\textit{i}}) an exact reduction; and, ({\textit{ii}}) an approximate reduction, which sets $\bN_i \equiv {\bf 0}$ and $\bM_i \equiv {\bf 0}$ and results in a block diagonal matrix $\hat\bS$. The solution approach adopted herein is based on ($ii$) and therefore each sub-system $\bR_i {\hat \bx}_i = {\hat \bg}_i$ is solved independently using the following steps:
\begin{subequations}
	\label{eq:solveTopBottom}
\begin{align}
\mbox{Form } &{\bar \bR}_i = {\bf I}_M - \T{\bW}{i+1} \B{\bV}{i} \\
\mbox{Solve } &{\bar \bR}_i \T{\tilde{\bx}}{i+1} = \T{\bg}{i+1} - \T{\bW}{i+1} \B{\bg}{i} \\
\mbox{Calculate } &\B{\tilde{\bx}}{i} = \B{\bg}{i} - \B{\bV}{i} \T{\tilde{\bx}}{i+1}
\end{align}
\end{subequations}
Note that a tilde was used to differentiate between the actual and approximate values $\T{\tilde{\bx}}{i}$ and $\B{\tilde{\bx}}{i}$ obtained upon dropping the $\bN_i$ and $\bM_i$ terms. An approximation of the solution of the original problem is finally obtained by solving independently and in parallel $P$ systems using the available LU factorizations of the $\bA_i$ matrices:
\begin{subequations}
\begin{alignat}{8}
% ----------- Purify x_1
&\:\bA_1 \bx_1 &
&= &
&\:\mathbf{b}_1 & 
&{} &
&{} &
&\:- &
&\left[\begin{matrix} \zero \\ \zero \\ \bB_1 \T{\tilde{\bx}}{2} \end{matrix}\right] &
&{} \\
% ----------- Purify x_i
&\:\bA_i \bx_i &
&=&
&\:\mathbf{b}_i & 
&-&
&\left[\begin{matrix} \bC_i \B{\tilde{\bx}}{i-1} \\ \zero \\ \zero \end{matrix}\right] &
&\:-&
&\left[\begin{matrix} \zero \\ \zero \\ \bB_i \T{\tilde{\bx}}{i+1} \end{matrix}\right] \, , &
&\quad i = 2,\ldots,P-1 \\
% ----------- Purify x_P
&\:\bA_P \bx_P &
&= &
&\:\mathbf{b}_P& 
&-&
&\left[\begin{matrix} \bC_P \B{\tilde{\bx}}{P-1} \\ \zero \\ \zero \end{matrix}\right] \; .&
&{} &
&{} &
&{}
\end{alignat}
\end{subequations}

Computational savings can be made by noting that if an LU factorization of the diagonal blocks $\bA_i$ is available, the bottom block of the right spike; i.e. 
$\B{\bV}{i}$, can be obtained from (\ref{eq:RightSpike_1}) using only the bottom $K \times K$ blocks of L and U. However, obtaining the top block of the left spike requires calculating the entire spike $\bW_i$. An effective alternative is to perform an additional UL factorization of $\bA_i$, in which case $\T{\bW}{i}$ can be obtained
using only the top $K \times K$ blocks of the new U and L.  

Next, note that the decision to set $\bN_i \equiv {\bf 0}$ and $\bM_i \equiv {\bf 0}$ relegates the resulting algorithm to preconditioner status. Embracing this path is justified by the following observation that although the dimension of the reduced linear system in (\ref{eq:SPIKE_reduced_system}) is smaller that that of the original problem, its half-bandwidth is at least three times larger. The memory footprint of exactly solving (\ref{eq:SPIKE_reduced_system}) is large, thus limiting the size of problems that can be tackled on the GPU. Specifically, at each recursive step, additional memory that is required to store the new reduced matrix cannot be deallocated until the global solution is fully recovered.

Finally, it becomes apparent that the quality of the preconditioner is correlated to neglecting the $\bN_i$ and $\bM_i$ terms. For the sake of this discussion, assume that the matrix $\bA$ is diagonally dominant with a degree of diagonal dominance $d \ge 1$; i.e.,
\begin{equation}
\label{eq:diagDominanceDef}
|a_{ii}| \ge d \sum\limits_{j \ne i} |a_{ij}| \; , \forall i = 1,\ldots,N \; .
\end{equation}
\noindent When $d>1$, the elements of the left spikes $\bW_i$ decay in magnitude from top to bottom, 
while those of the right spikes $\bV_i$ decay from bottom to top \cite{MiMa2008}.  This decay, which is more
pronounced the larger the degree of diagonal dominance of $\bA$, justifies the approximation $\bN_i \equiv {\bf 0}$ and $\bM_i \equiv {\bf 0}$. However, note that having $\bA$ be diagonal dominant, although desirable, it is not a prerequisite as demonstrated by numerical experiments reported herein. Truncating when $d<1$ will lead to a preconditioner of lesser quality.

\subsubsection{Nomenclature, solution strategies}
\label{sss:nomenclature}
Targeted for execution on the GPU, the methodology outlined above becomes the foundation of a parallel implementation called herein ``split and parallelize'' (SaP). The matrix $\bA$ is split into block diagonal matrices ${\bA}_i$, which are processed in parallel. The code implementing this strategy is called {\SpikeHyb}. Several flavors of {\SpikeHyb} can be envisioned. At one end of the spectrum, one solution path would implement the exact reduction, a strategy that is not considered herein. At the other end of the spectrum, {\SpikeHyb} solves the block-diagonal linear system in \ref{eq:diag-sys} and for preconditioning purposes uses the approximation ${\bf x} \approx {\bf g}$. In what follows, this will be called the decoupled approach, {\SaPD}. The middle ground is the approximate reduction, which sets $\bN_i \equiv {\bf 0}$ and $\bM_i \equiv {\bf 0}$. This will be called the coupled approach, {\SaPC}, owing to the coupling that occurs through the truncated spikes; i.e., $\B{\bV}{i}$ and $\T{\bW}{i+1}$.

Neither the coupled nor the decoupled paths qualify as direct solvers and {\SpikeHyb} employs an outer Krylov subspace scheme to solve $\bA \bx = \bb$.  The solver uses {\BCG}($\ell$) \cite{SlFo1993} and left-preconditioning, unless the matrix $\bA$ is symmetric and positive definite, in which case the outer loop implements a conjugate gradient method \cite{Saad2003}. {\SpikeHyb} is open source and available at \cite{SaP_git,SaPWebsite}.

\subsection{The sparse linear system case}
%\begin{remunerate}
%\item
%	Discuss overall algorithm -- this is a preconditioner that can be used with (almost any) iterative Krylov solver (0.5 pages, not much here)
%
%\item
%	Reorderings for boosting diagonal dominance (DB) and reducing bandwidth (CM) (0.5 pages for both of them). Essentially state what they are and why we need them.
%
%\item
%	Bring up second stage reordering: how done, and what implications it has (we do have to do full solves on the spikes). ({\em this is out first contribution -- this one on the algorithmic side}) (0.5 pages)
%\end{remunerate}

The discussion focuses next on solving $\bA_s \bx = \bb$, where ${\bA_s \in {\mathbb{R}}^{N \times N}}$ is assumed to be a sparse matrix. The salient attribute of the solution strategy is its fallback on the dense banded approach described in \S\ref{ss:denseLinSysExp}. Specifically, an aggressive row and column permutation process is employed to transform $\bA_s$ into a matrix $\bA$ that has a large $d$ and small $K$.  Although the reordered matrix will remain sparse within the band, it will be regarded to be dense banded and LU- and/or UL-factored accordingly.  For matrices $\bA_s$ that are either nonsymmetric or have low $d$, a first set of row permutations is applied as $\bQ \bA_s \bx = \bQ \mathbf{b}$, to either maximize the
number of nonzeros on the diagonal (maximum traversal search) \cite{Duff1981}, or maximize the product of the 
absolute values of the diagonal entries \cite{DuKo1999, DuKo2001}. Both reordering algorithms are implemented using 
a depth first search with a look-ahead technique similar to the one in the Harwell Software Library (HSL) \cite{HSL}. 

While the purpose of the first reordering $\bQ \bA_s$ is to render the permuted matrix diagonally ``heavy'', a second reordering seeks to reduce $K$ by using the traditional Cuthill-McKee {\CM} algorithm \cite{CuMc1969}. Since the diagonal entries should not be relocated, the second permutation is applied to the symmetric matrix $\bQ\bA_s + \bA_s^T\bQ^T$. Following these two reorderings, the resulting matrix $\bA$ is split to obtain ${\bA}_1$ through ${\bA}_P$. A third {\CM} reordering is then applied to each ${\bA}_i$ for further reduction of bandwidth. While straightforward to implement in {\SaPD}, this third stage reordering in {\SaPC} mandates computation of the entire spikes, an operation that can significantly increase the memory footprint and flop count of the numerical solution. Note that third stage reordering in {\SaPC} renders the UL factorization superfluous since computing only the top of a spike is insufficient. 

If $\bA_i$ is diagonally dominant, the LU and/or UL factorization can be safely carried out without pivoting \cite{golubMatrixBook96}. Adopting the strategy used in {\Pardiso} \cite{pardiso}, we always perform factorizations of the diagonal blocks $\bA_i$ {\emph{without}} pivoting but with {\em pivot boosting}. Specifically, if a pivot becomes smaller than a threshold value, it is boosted to a small, user controlled value $\epsilon$. This yields a factorization of a slightly perturbed diagonal block, $\bL_i \bU_i = \bA_i + \delta\bA_i$, where $\| \delta\bA_i \| = \mathcal{O}(u \| \bA\|)$ and $u$ is the unit roundoff~\cite{MSC2009}.

\subsubsection{Brief comments on the reordering algorithms}
\label{sss:reordering}
{\SpikeHyb} employs two reordering strategies, namely Diagonal Boosting ({\DB}) and Cuthill-McKee ({\CM}), possibly multiple times, to reduce $K$ and increase the degree of diagonal dominance. {\DB} is applied first at the matrix $\bA_s$ level, followed by {\CM} applied at matrix level, and possibly followed by a set of $P$ third-stage {\CM} reorderings applied at the sub-matrix $\bA_i$ level.

\noindent \textbf{Diagonal Boosting.} The {{\DB}} algorithm seeks to improve diagonal dominance in $\bA_s$ and draws on a minimum bipartite perfect matching~\cite{carpaneto1980algorithm,kuhn1955hungarian,burkhard1980assignment,carraresi1986efficient,derigs1986efficient,jonker1987shortest}. 
There are several variants of the algorithm aimed at different outcomes, e.g., maximizing the absolute value of bottleneck, the sum, the product or other metrics that factor in the diagonal entries. As a proxy for diagonal dominance, {\SpikeHyb} maximizes the absolute value of the product of all diagonal entries. 

The algorithm that seeks to leverage GPU computing is as follows. Given a matrix $\{a_{ij}\}_{n \times n}$, find a permutation $\sigma$ that maximizes $\prod_{i=1}^{n}|a_{i\sigma_{i}}|$. Denoting $a_i = \max_{j}|a_{ij}|$ and noting that $a_i$ is an invariant of $\sigma$, then we are to minimize
\[\log\prod\limits_{i=1}^{n}\frac{a_i}{|a_{i\sigma_{i}}|} = \sum\limits_{i=1}^n\log\frac{a_i}{|a_{i\sigma_{i}}|} = \sum\limits_{i=1}^{n}(\log a_i - \log |a_{i \sigma_{i}}|)\, .\]
The reordering problem is reduced to minimum bipartite perfect matching in the following way: given a bipartite graph $G_C = (V_R, V_C, E)$, we define the weight $c_{ij}$ of the edge between nodes $i \in V_R$ and $j \in V_C$ as
\begin{equation}
\label{eq:DBrelated}
c_{ij} = \begin{cases} \log a_i - \log |a_{ij}| & (a_{ij} \ne 0) \\ \infty & (a_{ij} = 0) \end{cases} \,.
\end{equation}
If we are able to find a minimum bipartite perfect matching $\sigma$ such that $\sum c_{i \sigma_{i}}$ is minimized, according to the process of reduction above, then $\prod_{i=1}^n |a_{i \sigma_i}|$ is maximized.

\noindent \textbf{Bandwidth reduction.} Whether ${\bf Q}\bA_s$ is sparse or not, there are $P-1$ pairs of always \emph{dense} spikes, each of dimension $N_i \times K$. They need to be stored unless one employs an LU and UL factorization of $\bA_i$ to retain only the appropriate bottom and top components. Large $K$ values pose memory challenges; i.e., storing and data movement, that limit the size of the problems that can be tackled. Moreover, the spikes need to be computed by solving multiple right-hand side linear systems with $\bA_i$ coefficient matrices. There are $2K$ such systems for each of the $P-1$ pairs of spikes. Evidently, a low $K$ is highly desirable. However, finding the lowest half-bandwidth $K$ by symmetrically reordering a sparse matrix is NP-hard. The {\CM} reordering provides simple and oftentimes effective heuristics to tackle this problem. Moreover, as the {\CM} reordering yields symmetric permutations, it will not displace the ``heavy'' diagonal terms obtained during the {\DB} step. However, to obtain a symmetric permutation, one has to start with a symmetric matrix. To this end, unless $\bA$ is already symmetric and does not call for a {\DB} step (which is the case, for instance, when $\bA$ is symmetric positive definite), the matrix passed over for {\CM} reordering is $(\bA + \bA^T)/2$. Given a symmetric $n \times n$ matrix with $m$ non-zero entries {\CM} works on its adjacency matrix. {\CM} first picks a random node and adds the node to the work list. Then the algorithm repeats sorting all its neighboring nodes with non-descending vertex degree and adding them until all vertices have been added and removed once from the work list. In other words, {\CM} is essentially a {\BFS} where neighboring vertices are visited in order from lowest to highest vertex degree. 

\noindent \textbf{Third-stage reordering.}
The {\DB}--{\CM} reordering sequence yields diagonally-heavy matrices of smaller bandwidth. The band itself however can be very sparse. The purpose of the third-stage {\CM} reordering is to further reduce the bandwidth within each $\bA_i$ and reduce the sparsity within the band. Consider, for instance, the matrix ANCF88950 that comes from structural dynamics \cite{serban2015}. It has \SI{513900}{} nonzeros, $N= 88\,950$, and an average of $5.78$ non-zero elements per row. After {\DB}--{\CM} reordering with no drop-off, the resulting banded matrix has a half-bandwidth $K = 205$. The band itself is very sparse with a fill-in of only $0.7\%$ within the band. In its default solution, {\SpikeHyb} constructs a block banded matrix where each diagonal block $\bA_i$, obtained after the initial {\DB}--{\CM} reorderings, is allowed to have a different bandwidth. 
This is achieved using another {\CM} pass, independently and in parallel for each $\bA_i$. Applying this strategy to ANCF88950, using $P = 16$ partitions, the half bandwidth is reduced for all partitions to values no higher than $K = 141$, while the fill-in within the band becomes approximately $3\%$. 

Note that this third-stage reordering does nothing to reduce the column-width of the spikes. However, it helps in two respects: a smaller memory footprint for the LU/UL factors, and less factorization effort. These are important side effects, since the LU/UL GPU factorization is currently done in-core considering $\bA_i$ to be \emph{dense} within the band.

%==============================================================================
\section{Brief implementation details}
\label{s:implementation}

\subsection{Dense banded matrix factorization details}
\label{ss:impl-details}
This subsection provides implementation details regarding how the $P$ partitions ${\bA}_i$ are determined, how the banded matrix $\bA$ is stored, and how the LU/UL steps are implemented on the GPU.

\vspace{2mm}

%----------------------------------------------------------------------------------------------------------------------------------------------
% Load balancing
%----------------------------------------------------------------------------------------------------------------------------------------------
\noindent\textbf{Number of partitions and partition size.}
The selection of $P$ must strike a balance between two conflicting requirements. On the one hand, having a large $P$ is attractive given that the LU/UL factorization of $\bA_i$ for $i=1,\ldots,P$ can be done independently and simultaneously. On the other hand, this negatively impacts the quality of the resulting preconditioner, due to the approximations in evaluating the spikes corresponding to the coupling of the diagonal blocks $\bA_{i}$ and $\bA_{i+1}$. Since this adversely impacts the quality of the resulting preconditioner, a high $P$ could lead to poor preconditioning and an increase in the number of iterations to convergence. In the current implementation, no attempt is made to automate this selection and some experimentation is required. 

Given a $P$ value, the size of the diagonal blocks $\bA_i$ is selected to achieve load balancing. The first $P_r$ partitions are of size $\lfloor N / P \rfloor + 1$, while the remaining are of size $\lfloor N / P \rfloor$, where $N = P \lfloor N / P \rfloor + P_r$.

\vspace{2mm}

%----------------------------------------------------------------------------------------------------------------------------------------------
% Matrix storage
%----------------------------------------------------------------------------------------------------------------------------------------------
\noindent\textbf{Matrix storage.}
For general dense banded matrices $\bA_i$, we adopt a ``tall and thin'' storage in column-major order. All diagonal elements are stored in the $K$-th column. The rest of the elements are correspondingly distributed columnwise. This strategy, shown below for a matrix with $N=8$ and $K=2$, groups the operands of the LU/UL factorizations and allows coalesced memory accesses that can fully leverage the GPU's bandwidth.
\[\begin{bmatrix}
	*          & *          & a_{11} & a_{21} & a_{31} \\
	*          & a_{12} & a_{22} & a_{32} & a_{42} \\
	a_{13} & a_{23} & a_{33} & a_{43} & a_{53} \\
	a_{24} & a_{34} & a_{44} & a_{54} & a_{64} \\
	a_{35} & a_{45} & a_{55} & a_{65} & a_{75} \\
	a_{46} & a_{56} & a_{66} & a_{76} & a_{86} \\
	a_{57} & a_{67} & a_{77} & a_{87} & *      \\
	a_{68} & a_{78} & a_{88} & *      & *      \\
\end{bmatrix}\]

\vspace{2mm}

%----------------------------------------------------------------------------------------------------------------------------------------------
% LU/UL/Cholesky implementation
%----------------------------------------------------------------------------------------------------------------------------------------------
\noindent\textbf{LU/UL factorizations.}
The solution strategy pursued calls for an LU and an optional UL factorization of each dense banded diagonal block $\bA_i$. The implementation requires a certain level of synchronization since for each $\bA_i$, the factorization, forward elimination, and backward substitution phases each consist of $N_i - 1$ dependent steps that need to be choreographed.  One aggravating factor is the GPU lack of native, low overhead, support for synchronization between threads running in different blocks. The established GPU strategy for inter-block synchronization is ``exit and launch a new kernel''. This guarantees synchronization at the GPU-grid level at the cost of non-negligible overhead. In a trade-off between minimizing the overhead of kernel launches and maximizing the occupancy of the GPU, we established two execution paths: one for $K < 64$, the second one for larger bandwidths. As a side note, the threshold value of $64$ was selected through numerical experimentation over a variety of problems and is controlled by the number of threads that can be organized in a block in CUDA \cite{cudaProgGuide}.

For $K < 64$, the code was designed to reduce the kernel launch count.
Instead of having $N_i-1$ kernel launches, each completing a step of the factorization of $\bA_i = {\bf L}_i {\bf U}_i$ by updating entries in a $(K+1)\times(K+1)$ window of elements, a single kernel is launched to factor $\bA_i$. It uses $\min(K^2, 1024)$ threads per block and relies on low-overhead stream-multiprocessor synchronization support \emph{within} the block, without any need for global synchronization.
In a so-called {\em window-sliding} method, at each step of the factorization; i.e., during the process of computing column entries in L and row entries of U, each thread updates a fixed number of $\bA_i$ entries. On current GPU hardware, this fixed number is between $1$ and $4$.
Once all threads in the block complete their work, they are synchronized and the $(K+1)\times(K+1)$ window slides down by one row and to the right by one column. The value $4$ is explained as follows. Assume that $K=63$. Then, 
the sliding window has size $64 \times 64$. Since the two-dimensional GPU thread block size is $1024=32\times32$, each thread will handle four entries of the window of focus.

For $K \ge 64$, {\SaP} uses multiple blocks of threads to update L and U entries. On the upside, there are more threads working on the window of focus. On the downside, there is overhead associated with leaving and reentering the kernel, a process that has the side effect of flushing the shared memory and registers. The window is larger than $K \times K$, and it slides at a stride of eight; i.e., moves down by eight rows and to the right by eight columns upon exiting and reentering the LU factorization kernel. 

%----------------------------------------------------------------------------------------------------------------------------------------------
% Use of registers and shared memory
%----------------------------------------------------------------------------------------------------------------------------------------------
\noindent\textbf{Use of registers and shared memory.}
If the user decides to employ a third-stage reordering, the coupling sub-blocks $\bB_i$ and $\bC_i$ are used to compute the entire spikes in a scheme that renders a UL factorization superfluous.
Then, $\bB_i$ and $\bC_i$ are each first partitioned into sub-blocks of dimension $L \times K$ where $L$ is at most $20$.
Each forward/backward sweep to get the spikes is unrolled, and in each iteration of the new loop, one entire sub-block, rather than
a vector of length $K$, is calculated.
To this end, the corresponding elements in the matrix $\bA_i$ are pre-fetched into shared memory and the entries of the sub-block are preloaded into registers.
This strategy, in which all operations to calculate the spikes draw on registers and shared memory, leads to 50\% to 70\% improvement in performance when compared with an alternative that calculates the spike elements in a loop without leveraging the low latency/high bandwidth of the GPU register file and shared memory.

%----------------------------------------------------------------------------------------------------------------------------------------------
% Mixed Precision
%----------------------------------------------------------------------------------------------------------------------------------------------
\noindent\textbf{Mixed Precision Strategy.}
The solution uses a mixed-precision implementation by falling back on single precision for the preconditioner and switching to double precision arithmetic in the outer {\BCGtwo} calculations. A battery of tests indicate that
this strategy results in a 50\% average reduction in time to solution when compared with an approach where all calculations are performed in double precision.

%-------------------------------------------------------------------------------
\subsection{{\DB} reordering implementation details}
\label{ss:db-impl}
{\SpikeHyb} organizes the {\DB} algorithm into four stages, {\DBS{1}} through {\DBS{4}}. Due to differences in the nature and degree of parallelism of these stages, {\DB} implements a hybrid strategy; namely, it relies on GPU computing for {\DBS{1}} and {\DBS{4}} and on CPU computing for {\DBS{2}} and {\DBS{3}}. A thorough discussion of the implementation is provided in~\cite{AngMC64-2014}. Therein, a solution that kept the entire {\DB} implementation on the GPU was discussed and deemed decisively slower than the hybrid strategy adopted here.

\vspace{0.5mm}	

\noindent {\bf {\DBS{1}}: form bipartite graph.} This stage assembles a matrix that mirrors the structure of the original sparse matrix. The sparsity pattern of the input matrix is maintained and the values of its nonzero entries are modified according to Eq.~(\ref{eq:DBrelated}). The stage is highly parallel and involves: (1) calculating for each row of the original matrix the max absolute value, and (2) updating each value to form the weighted bipartite graph.

\vspace{0.5mm}	

\noindent {\bf {\DBS{2}}: find initial partial match.} This stage is not mandatory but the availability of an initial partial match as a starting point for the next stage was found to considerably reduce the running time for the overall algorithm \cite{AngMC64-2014}. Like in~\cite{carpaneto1980algorithm}, after setting $u_i = \min_j c_{ij}$ and $v_j = \min_i(c_{ij} - u_i)$, we try to match as many pairs of nodes as possible. The matched nodes $(i, j)$ should satisfy $u_i + v_j = c_{ij}$. This yields augmenting paths of length one. This stage, which was implemented to execute in parallel, was compute intensive as it had to resolve scenarios where multiple column nodes would match the same row node. A CPU parallel implementation was found to be more suitable owing to intense integer arithmetic and control flow overhead.	

\vspace{0.5mm}	

\noindent {\bf {\DBS{3}}: find perfect match.} Finding matches in a bipartite graph $G_C$ is equivalent to finding the shortest paths in an associated reduced graph. Omitting some of the details, the shortest path problem is tackled using Dijkstra's algorithm~\cite{dijkstra1959note}, which is applied to all nodes $i$ that are unmatched in the initial partial match obtained in {\bf {\DBS{2}}}. This ensures that all row nodes, and therefore all column nodes, are eventually matched.  The theoretical complexity of this stage is $O(n \cdot (m+n) \cdot \log n)$, where $n$ and $m$ are the dimension and number of nonzeros in the input matrix, respectively. However, thanks to the preprocessing {\bf {\DBS{2}}}, actual run times for finding a perfect match are acceptable in all situations and this stage is the {\DB} bottleneck only for about half of the matrices tested~\cite{AngMC64-2014}.

\vspace{0.5mm}	

\noindent {\bf {\DBS{4}}: extract permutation and scaling factors.} The matrix permutation can be obtained directly from the resulting perfect match: if the row node $i$ was matched to the column node $j$ then rows (or columns) $i$ and $j$ must be permuted. Optionally, scaling factors can be calculated and applied to rows and columns in order to bring the matrix to a so-called $I$-matrix form; i.e., a matrix with $1$ or $-1$ on the diagonal and off-diagonal elements of absolute value less than $1$, see~\cite{olschowka1996new}.  This stage is highly parallelizable and amenable to GPU computing.

%-------------------------------------------------------------------------------
\subsection{{\CM} reordering implementation details}
\label{ss:cm-impl}
The unordered {\CM} algorithm, which draws on an approach described in~\cite{karantasis2014parallelization}, is separated into three stages, {\CMS{1}} through {\CMS{3}}.
A high quality reordering calls for several {\BFS} iterations, which are called herein ``{\CM} iterations''. Just like the {\DB} implementation, the {\CM} solution ($i$) is hybrid -- the overall algorithm leverages both CPU and GPU computing; and, ($ii$) it uses CPU--GPU unified memory, a recent CUDA feature \cite{unifiedMemoryCUDA}, to provide for a simple and transparent memory management process. The latter feature allows the CUDA runtime to transparently manage the CPU--GPU data migration as the computation switches back and forth between the CPU and GPU. Since no explicit, programmer initiated, data transfer is required, the code is cleaner and more concise.

% ------------------------------------------------------
\noindent {\bf {\CMS{1}}: pre-processing.} The first stage is implemented on the GPU to accomplish two objectives. First, it produces the data structure that is worked upon.
As the input matrix $\bA$ is not guaranteed to be symmetric, the sparse matrix structure for $(\bA + \bA^T)/{2}$ is produced in anticipation of the subsequent two stages of the algorithm. Second, in order to avoid repetitively sorting the neighbors of a given node, the nodes with the same row indices are pre-sorted by ascending vertex degree of column index.

\vspace{1mm}

\noindent {\bf {\CMS{2}}: perform standard BFS.} After experimenting with the implementation, the strategy adopted started from several nodes and in parallel performed what would be a traditional {\CMS{2}} \& {\CMS{3}} combo. 
The alternative of considering one node only, namely the node with the smallest vertex degree, yields a second level {\BFS} tree with fewer nodes. Eventually, the resulting {\BFS} tree will likely be ``tall and thin''. Starting from several nodes and completing the reordering process for each of them increases the likelihood of avoiding a ``bad'' initial node. In practical terms, owing to the use of parallel computing, this strategy yields smaller bandwidths at a modest increase in computational overhead.

For each starting node, a standard {\BFS} pass yields the levels of all nodes in the {\BFS} tree. Since the order of nodes at the same level is not critical in this stage, parallel computing can help by concurrently visiting the neighbors of all nodes at the previous level. We use an outer loop to iterate over the levels, and in each iteration, depending on the number of nodes $n_p$ added in the previous iteration, we decide whether this iteration is executed on the GPU or CPU. The heuristics used are as follows: a kernel handles the iteration on the GPU only if $n_p \ge 10$. There are two notable implementation details. First, the {\CM} iterations are executed sequentially. After each iteration, we select the node at the previous level with the lowest vertex degree which has not yet been selected yet. If no such nodes exist; i.e., all nodes at the last level have been selected as starting nodes in previous iterations, a random node which has not been considered is selected. Second, the {\CM} iterations terminate either when the height of the {\BFS} tree does not increase, or when the maximum number of nodes over all levels does not decrease compared with the candidate optimal found so far. This strategy is proposed in ~\cite{PaduaHaiek:1980:MDT:909338} with the caveat that we only consider the leaf with the minimum degree. From practical experience, these heuristics lead to an algorithm that for most matrices terminates within three {\CM} iterations.

\vspace{1mm}

\noindent {\bf {\CMS{3}}: reorder nodes.}
The previous stage determines the level of each node. Roughly speaking, nodes are ordered in ascending order, from level 0 up to the maximum level $m_l$ and memory space can be pre-allocated for nodes at each level. Parallel computing is leveraged by observing that the order of nodes at level $l$ depends only on the order of nodes at level $l-1$. To that end, a pair of read/write pointers is set for each level, and except for level $0$, the read/write pointers of each level will point to the starting position of the level's pre-allocated space. We say a thread ``works on'' level $l$ if it reads nodes at level $l$ and writes their neighbors that are at level $l + 1$. Thus the execution thread working on level $l$ will read and modify the read pointer of level $l$ and the write pointer of level $l + 1$, and it will only read the write pointer of level $l$. Once the thread finishes reading all nodes at level $l$, it moves on to another level; otherwise it repeats checking whether or not the thread working on level $l-1$ has written nodes which it has not processed by checking if the read pointer at level $l$ lags the write pointer at level $l$. If yes, the thread working on level $l$ processes these nodes, i.e., writes their neighbors with level $l + 1$, and goes back to checking again whether it has finished processing or not; otherwise, it spins and waits for the thread working on the previous level. Note that the parallelism in {\CMS{3}} is rather coarse-grained and proved to be better suited for execution on the CPU.

%-------------------------------------------------------------------------------
\subsection{{\SpikeHyb}--components and computational flow}
\label{ss:overallAlg}
%----------------------------------------------------------------------------------------------------------------------------------------------
% Overall flowchard. Use this to later talk about profiling of the approach.
%----------------------------------------------------------------------------------------------------------------------------------------------
In the absence of column/row reordering before the LU factorization and pivoting during the factorization, the {\SpikeHyb} dense banded linear system solver is straightforward to implement. Upon partitioning $\bA$ into diagonal blocks $\bA_i$, each $\bA_i$ is subject to an LU factorization that requires an amount of time $T_{LU}$. Next, in $T_{BC}$ time, the coupling block matrices $\bB_i$ and $\bC_i$ are extracted on the GPU. The $\bV_i$ and $\bW_i$ spikes are subsequently computed in an operation that requires $T_{SPK}$ time. Afterwards, in $T_{LUrdcd}$ time, the spikes are truncated and the steps outlined in Eq.~(\ref{eq:solveTopBottom}) are taken to produce the intermediary values $\T{\tilde{\bx}}{i}$ and $\B{\tilde{\bx}}{i}$. At this point, the pre-processing step is over and two sets of factorizations, for $\bA_i$ and ${\bar \bR}_i$, are available for preconditioning during the iterative phase of the solution. The amount of time spent iterating is $T_{Kry}$, the iterative methods considered being {\BCGtwo} and conjugate gradient.

The sparse linear system solution is slightly more convoluted at the front end. A sequence of two permutations, {\DB} requiring $T_{DB}$ and {\CM} requiring $T_{CM}$ time, are carried out to increase the size of the diagonal elements and reduce bandwidth, respectively. An additional amount of time $T_{Drop}$ might be spent to drop off-diagonal elements in order to decrease the bandwidth of the reordered $\bA$ matrix. Since the {\DB} and {\CM} reorderings are hybrid, $T_{Dtransf}$ is used to keep track of the overhead associated with moving data back and forth between the CPU and GPU during the reordering process. An amount of time $T_{Asmbl}$ is spent on the GPU in book-keeping required to turn the reordered sparse matrix into a dense banded matrix. 

% -----------------
\begin{figure}[ht]
	\centering {\includegraphics[width=0.95\textwidth]{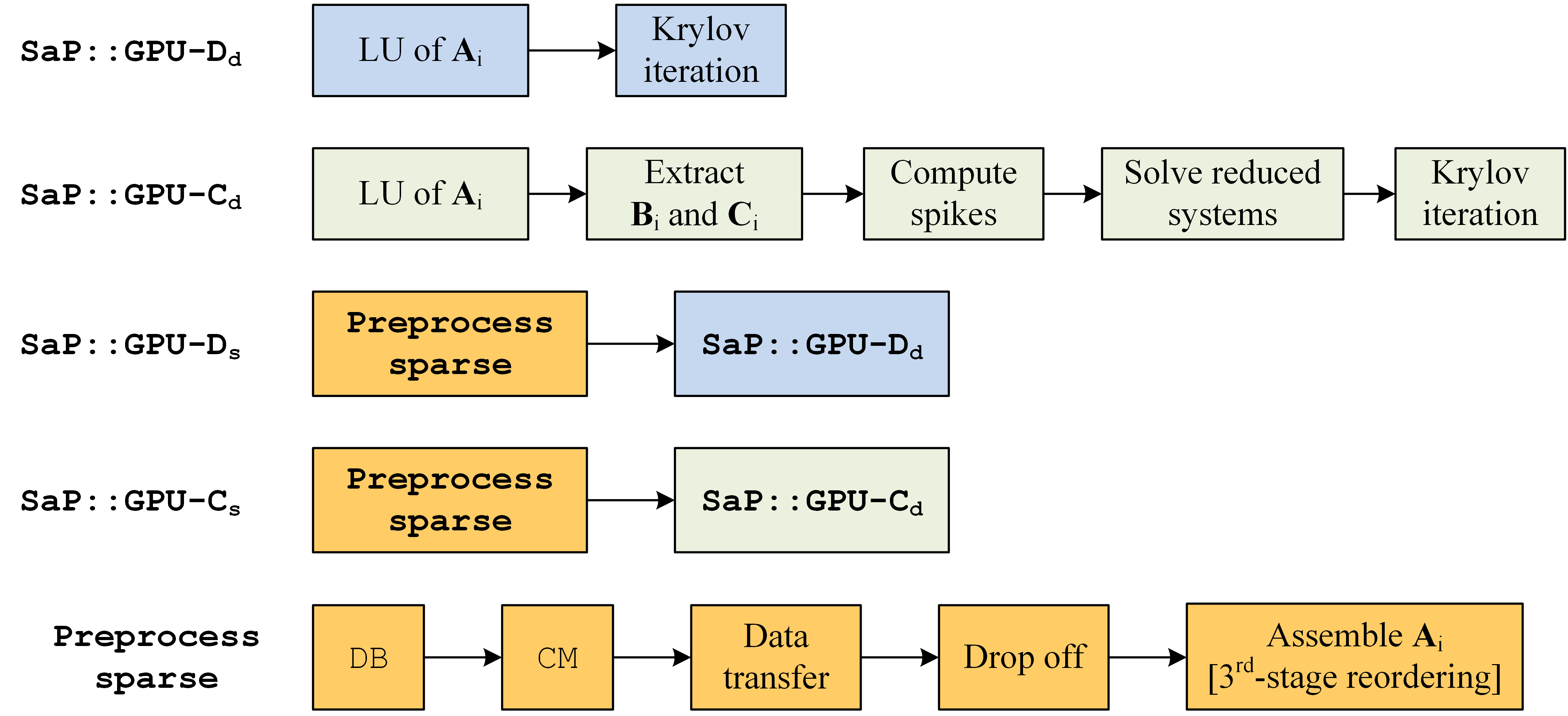}}
	\caption{Computational flow for {\SpikeHyb}.}
	\label{f:comFlowSpike}
\end{figure}
% -----------------

The process described above is summarized in Fig. \ref{f:comFlowSpike}. The boxes in gray are associated with the solution of a dense banded linear system. For a sparse linear system solve that uses a coupled approach; i.e., {\SaPC}, the total time is $T_{TotSparse}=T_{PrepSp}+T_{TotDense}$, where $T_{PrepSp}=T_{DB}+T_{CM}+T_{Dtransf}+ T_{Drop}+T_{Asmbl}$ and $T_{TotDense}=T_{LU}+T_{BC}+T_{SPK}+T_{LUrdcd}+T_{Kry}$. 
For {\SaPD}, owing to the decoupled nature of the solution, $T_{TotDense} = T_{LU}+T_{Kry}$, where $T_{LU}$ includes an {\CM} process that reduces the bandwidth of each $\bA_i$.
The names introduced; i.e., $T_{DB}$, $T_{CM}$, $T_{LUrdcd}$, etc., are referenced in the profiling study discussed in \S\ref{sss:profilingStudy} and used \emph{ad verbum} on the {\SpikeHyb} web-page \cite{SaPWebsite} to report profiling results for approximately 120 linear systems.

%\begin{itemize}
%	\item MC64: time to run MC64 reordering for the matrix on the CPU 
%	\item RCM: time to run RCM reordering for the matrix on the CPU 
%	\item Drop: time to drop off off-diagonal elements to decrease bandwidth. Done on the CPU. 
%	\item Dtransf: data transfer from CPU to GPU 
%	\item Asmbl: after reordering and drop-off, copy the sparse matrix to banded matrix stored in CPU memory 
%	\item BC: time required to get off-diagonal right hand sides (Bs and Cs) from the banded matrix - done on the GPU 
%	\item LU: LU time 
%	\item SPK: time to solve for the spikes Vs and Ws - done on the GPU 
%	\item LUrdcd: time required to factorize the reduced matrices - done on the GPU 
%	\item Kry: time spent in the Krylov solver (on the GPU) 
%\end{itemize}

%==============================================================================
\section{Numerical Experiments}
\label{s:experiments}
The next three subsections summarize results from three numerical experiments concerned, in this order, with the solution of dense banded linear systems, sparse matrix reordering, and the solution of sparse linear systems. The subsection order is meant to emphasize that dense banded linear system solution and matrix reordering are two prerequisites for an effective sparse linear system implementation in {\SpikeHyb}. The hardware/software setup for these numerical experiments is as follows. The GPU used was Tesla K20X \cite{TeslaK20,TeslaK20-datasheet}. {\SpikeHyb} uses {\texttt{CUDA 7.0}} \cite{cudaProgGuide}, {\texttt{cusp}} \cite{Cusp2012}, and {\texttt{Thrust}} \cite{thrust}. The CPU used was the 3GHz, 25 MB last level cache, Intel Xeon E5-2690v2. The node used hosted two such CPUs, which is the maximum possible for this type of chip, for a total of 20 cores executing up to 40 HTT threads. The two-CPU node was used to run Intel's MKL version 13.0.1, {\Pardiso} \cite{schenk2004solving}, {\MUMPS} \cite{mumps}, {\SuperLU} \cite{demmel2011superlu}, and Harwell's {\MCs} and {\MCsf} \cite{HSL}. Unless otherwise stated, all times reported are in seconds and were obtained on a dedicated machine. In an attempt to avoid warm up overhead, the results reported represent averages that drew on multiple successive identical runs.

When reporting below the results of several numerical experiments, one legitimate question is whether it makes sense to compare performance results obtained on one GPU with results obtained on two multicore CPUs. The multicore CPU is not the fastest, as Intel chips with more cores are presently available. Additionally, the Intel chip's microarchitecture is not Haswell, which is more recent than the Ivy Bridge microarchitecture of the Xeon E5-2690v2. Likewise, on the GPU side, one could have used a Tesla K80 card, which has roughly four times more memory than K20x and twice its memory bandwidth. Moreover, price-wise, the K80 would have been closer to the cost of two CPUs than K20x is. Finally, Kepler is not the latest microarchitecture either, since Maxwell currently enjoys that status. We do not attempt to answer these questions and hope that the interested reader will modulate this study's conclusions by factoring in unavoidable CPU--GPU hardware differences. No claim is made herein of one architecture being superior since such a claim could be easily proved wrong by moving from algorithm to algorithm or from discipline to discipline. The sole and narrow purpose of this section is to report on how apt {\SpikeHyb} is in tackling linear algebra tasks. To that end its performance is compared to that of established solutions running on CPUs and also of a recent GPU library.

\subsection{Numerical experiments related to dense banded linear systems}
\label{ss:theDBcase}
The discussion in this subsection draws on a subset of results reported in \cite{AngSPIKEMKL-2014} and presents results pertaining to the influence on {\SaP}'s time to solution of the number of partitions $P$ and of the diagonal dominance $d$ of the coefficient matrix, as well as a comparison against Intel's MKL solver over a spectrum of problem dimensions $N$ and half bandwidth values $K$. 

\subsubsection{Sensitivity with respect to $P$}
\label{sss:sensitivityWRT-P}
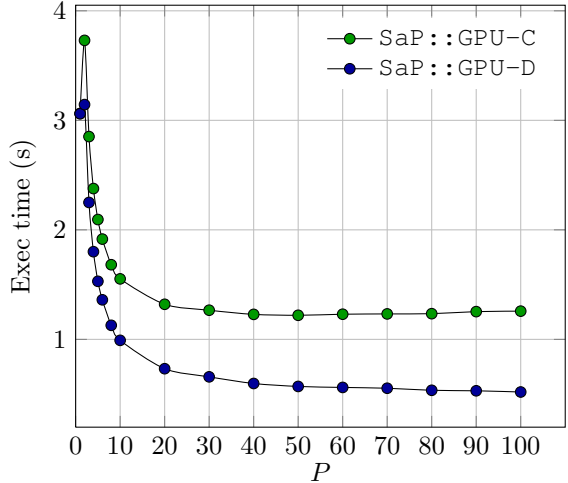
\begin{figure}
\centering
	\begin{tikzpicture}
	\begin{axis} [
%	at=(top.below south west),
	anchor=north west,
	scale only axis,
	width=0.5\textwidth, 
	xmajorgrids=true, ymajorgrids=true,
	xtick={0,10,20,30,40,50,60,70,80,90,100},
	xmin=0,
	xlabel style={yshift=0.5em},
	xlabel=$P$, 
	ylabel style={yshift=-1.5em},
	ylabel=Exec time (s),
	legend pos=north east, legend style={draw=none}, legend cell align=left
	]
	\addplot[styleA] table[x=P,y expr=\thisrow{SPIKETime}/1000]{Data/Banded_P.dat};
	\addlegendentry{{\SaPC}}
	\addplot[styleA2] table[x=P,y expr=\thisrow{BBDTime}/1000]{Data/Banded_P.dat};
	\addlegendentry{{\SaPD}}
	\end{axis}
	\end{tikzpicture}
\caption{Time to solution as a function of the number of partitions $P$. Study carried out for a dense banded linear system with $N=\SI{200000}{}$, $K=200$, and $d=1$.}
\label{f:P-sweep}
\end{figure}

The entire {\SpikeHyb} solution for dense banded linear systems is implemented on the GPU. We first carried out a sensitivity analysis of the time to solution with respect to the number of partitions. The results are summarized in Fig. \ref{f:P-sweep}. This behavior; i.e., relatively small gains after a threshold value of $P$, is typical. As a rule of thumb, some experimentation is necessary to find an optimal $P$ value. Otherwise, a conservatively large value should be picked in the neighborhood of 50 or above. For {\SaPD}, larger values of $P$ help with load balancing, particularly for GPUs with many stream multiprocessors. The same argument can be made for {\SaPC}, with the caveat that the spike truncation factor comes into play in a fashion that is modulated by the value of $d$.

% -----------------
% -----------------

It is instructive to see how the solution time is spent by {\SaPC} and {\SaPD} and understand how changing $P$ influences this distribution of the time to solution between the major implementation components. The results in Table \ref{t:sweepP4DvsCdense} provide this information as they compare the coupled and decoupled strategies in regards to the factorization times, $D_{pre}$ vs. $C_{pre}$; number of iterations in the Krylov solver, $D_{it}$ vs. $C_{it}$; amount of time spent iterating to find the solution at a level of accuracy of at least $10^{-10}$, $D_{Kry}$ vs. $C_{Kry}$; and the total times, $D_{Tot}$ vs. $C_{Tot}$. These times are defined as $D_{pre} = T_{LU}$, $C_{pre}=T_{LU}+T_{BC}+T_{SPK}+T_{LUrdcd}$, $D_{Tot} = D_{pre} + D_{Kry}$, and $C_{Tot}= C_{pre} + C_{Kry}$. Note that for {\SpikeHyb}, quarters of number of iterations are reported. This is due to the fact that {\BCGtwo} contains three exits points during each iteration. Moving from one to the next roughly requires the same amount of effort, which justifies the adopted convention.

The number of iterations to convergence suggests that the quality of the coupled-version of the preconditioner is superior. Yet the price for getting this better preconditioner is higher and {\SaPD} ends up winning by taking as little as half the time required by {\SaPC}. When the same factorization is used multiple times, this conclusion could change since the metric that controls the performance would be $D_{Kry}$ and $C_{Kry}$, or its number of iterations for convergence proxy. Also note that the return on increasing the number of partitions gradually fades away and for the coupled strategy there is no reason to go beyond $P=50$.

\pgfplotstableset{
	columns={P,T-PreP-D,T-PreP-C,nItrs-D,nItrs-C,T-Kry-D,T-Kry-C,T-Total-D,T-Total-C,SpeedUp},
	columns/P/.style={
		column name=$P$,
		int detect},
	columns/T-PreP-D/.style={
		dec sep align,
		column name=$D_{pre}$,
		fixed,
		precision=1},
	columns/T-PreP-C/.style={
		dec sep align,
		column name=$C_{pre}$,
		fixed,
		precision=1},
	columns/nItrs-D/.style={
		column name=${D}_{it}$,
		fixed,
		precision=2},
	columns/nItrs-C/.style={
		column name=${C}_{it}$,
		fixed,
		precision=2},
	columns/T-Kry-D/.style={
		dec sep align,
		column name=$D_{Kry}$,
		fixed,
		precision=1},
	columns/T-Kry-C/.style={
		dec sep align,
		column name=$C_{Kry}$,
		fixed,
		precision=1},
	columns/T-Total-D/.style={
		dec sep align,
		column name=$D_{Tot}$,
		fixed,
		precision=1},
	columns/T-Total-C/.style={
		dec sep align,
		column name=$C_{Tot}$,
		fixed,
		precision=1},
	columns/SpeedUp/.style={
		dec sep align,
		column name=\textsc{SpdUp},
		fixed,
		precision=2},
	every head row/.style={
		before row=\toprule,after row=\midrule},
	every last row/.style={
		after row=\bottomrule}}

\begin{table} [ht]
	\footnotesize
	\centering
\pgfplotstabletypesetfile{Data/sweepP4DvsCdense.txt}
\caption{Performance comparison over a spectrum of number of partitions $P$ for coupled (C) vs. decoupled (D) strategies in {\SpikeHyb}. All timings are in milliseconds. Problem parameters: $N= \SI{200000}{}$, $d=1$, $K=200$. The symbols used are as follows: $D_{pre}$--amount of time spent in preprocessing by the decoupled strategy; ${D}_{it}$--number of Krylov iterations for convergence; $D_{Tot}$--amount of time to converge. Similar values are reported for the coupled scenario. \textsc{SpdUp}$=D_{Tot}/C_{Tot}$.} 
\label{t:sweepP4DvsCdense}
\end{table}

\subsubsection{Sensitivity with respect to $d$}
\label{sss:sensitivityWRT-d}
\begin{figure}[ht]
\centering
	\begin{tikzpicture}
	\begin{axis}[
		scale only axis,
		width=0.6\textwidth, height=0.4\textwidth,
		xmajorgrids=true, ymajorgrids=true,
		ymax=6.5,
		xtick={0.1,0.3,0.5,0.7,0.9,1.1},
		ylabel style={yshift=-1.5em},
		ylabel=Speedup,
		name=top,
		legend pos=north east, legend style={draw=none}, legend cell align=left
		]
		\addplot[styleA] table[x=D,y expr=\thisrow{MKLTime}/\thisrow{T50}]{Data/Banded_D.dat};
		\addlegendentry{{\SaPC} over {\MKL}}
		\addplot[styleA2] table[x=D,y expr=\thisrow{MKLTime}/\thisrow{bT50}]{Data/Banded_D.dat};
		\addlegendentry{{\SaPD} over {\MKL}}
	\end{axis}

	\begin{axis} [
		at=(top.below south west),
		anchor=north west,
		scale only axis,
		width=0.6\textwidth, 
		xmajorgrids=true, ymajorgrids=true,
		xtick={0.1,0.3,0.5,0.7,0.9,1.1},
		ymax=4,
		xlabel style={yshift=0.5em},
		xlabel=$d$, 
		ylabel style={yshift=-1.5em},
		ylabel=Exec time (s),
		legend pos=north east, legend style={draw=none}, legend cell align=left
		]
		\addplot[styleA] table[x=D,y expr=\thisrow{T50}/1000]{Data/Banded_D.dat};
		\addlegendentry{{\SaPC} }
		\addplot[styleA2] table[x=D,y expr=\thisrow{bT50}/1000]{Data/Banded_D.dat};
		\addlegendentry{{\SaPD} }
		\addplot[styleB] table[x=D,y expr=\thisrow{MKLTime}/1000]{Data/Banded_D.dat};
		\addlegendentry{{\MKL}}
	\end{axis}
	\end{tikzpicture}
\caption{Influence of the diagonal dominance $d$, with $0.06 \le d \le 1.2$, for fixed values $N=\SI{200000}{}$, $K=200$ and $P = 50$.}
\label{f:SPIKE_MKL_banded:d}
\end{figure}
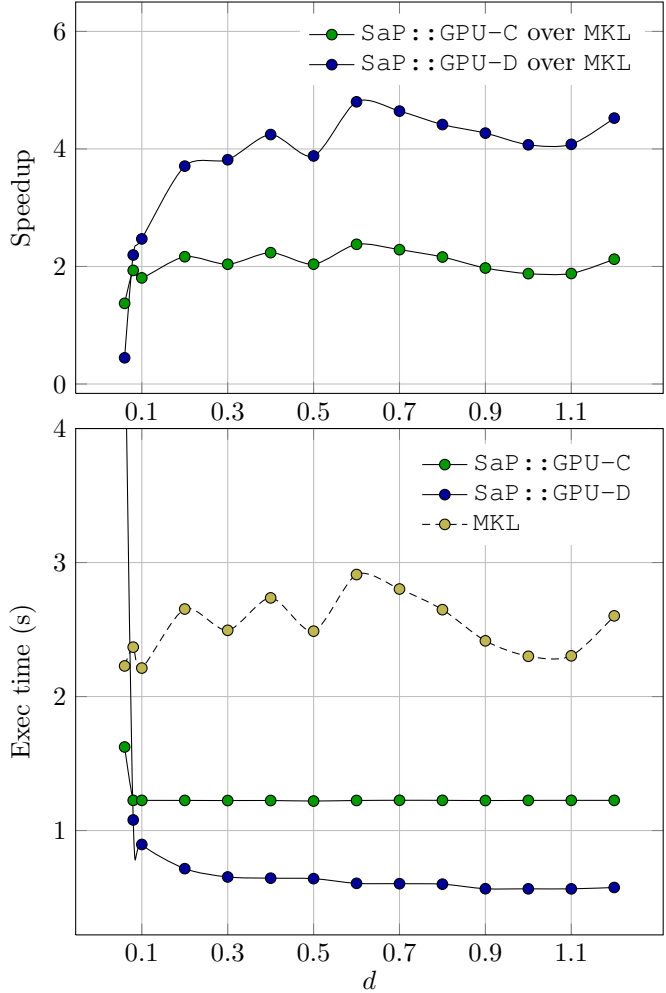

Next, we report on the performance of {\SpikeHyb} for a dense banded linear system with $N=\SI{200000}{}$ and $K=200$, for degrees of diagonal dominance in the range $0.06 \le d \le 1.2$, see Eq.~(\ref{eq:diagDominanceDef}). The entries in the matrix are randomly generated and $P=50$.
The findings are summarized in Fig.~\ref{f:SPIKE_MKL_banded:d}, where {\SaPC} and {\SaPD} are compared against the banded linear solver in {\MKL}. When $d>1$ the impact of the truncation becomes increasingly irrelevant, a situation that places the {\SpikeHyb} at an advantage. As such, there is no reason to go beyond $d=1.2$ since if anything, the results will get better. The more interesting range is $d<1$, when the diagonal dominance requirement is violated.
{\SpikeHyb} solver demonstrates uniform performance over a wide range of degrees of diagonal dominance. For instance, {\SaPC} typically required less than one Krylov iteration for all $d > 0.08$. As the degree of diagonal dominance decreases further, the number of iterations and hence the time to solution increase significantly as a consequence of truncating the spikes that now contain non-negligible values.

It is instructive to see how the solution time is spent by {\SaPC} and {\SaPD} and understand how changing $d$ influences this distribution of the time to solution between the major implementation components. The results reported in Table \ref{t:sweepd4DvsCdense} provide this information as they help answer the following question: can one still use a decoupled approach for matrices that are far from being diagonal dominant? The answer is yes, except in the most extreme case, when $d=0.06$. Note that the number of iterations to convergence for the decoupled approach quickly recovers away from small values of $d$. In the end, the same $2\times$ speedup factor is obtained virtually over the entire spectrum of $d$ values.

\pgfplotstableset{
	columns={d,T-PreP-D,T-PreP-C,nItrs-D,nItrs-C,T-Kry-D,T-Kry-C,T-Total-D,T-Total-C,SpeedUp},
	columns/d/.style={
		dec sep align,
		column name=$d$,
		precision=2},
	columns/T-PreP-D/.style={
		dec sep align,
		column name=$D_{pre}$,
		fixed,
		precision=1},
	columns/T-PreP-C/.style={
		dec sep align,
		column name=$C_{pre}$,
		fixed,
		precision=1},
	columns/nItrs-D/.style={
		column name=${D}_{it}$,
		fixed,
		precision=2},
	columns/nItrs-C/.style={
		column name=${C}_{it}$,
		fixed,
		precision=2},
	columns/T-Kry-D/.style={
		dec sep align,
		column name=$D_{Kry}$,
		fixed,
		precision=1},
	columns/T-Kry-C/.style={
		dec sep align,
		column name=$C_{Kry}$,
		fixed,
		precision=1},
	columns/T-Total-D/.style={
		dec sep align,
		column name=$D_{Tot}$,
		fixed,
		precision=1},
	columns/T-Total-C/.style={
		dec sep align,
		column name=$C_{Tot}$,
		fixed,
		precision=1},
	columns/SpeedUp/.style={
		dec sep align,
		column name=\textsc{SpdUp},
		fixed,
		precision=2},
	every head row/.style={
		before row=\toprule,after row=\midrule},
	every last row/.style={
		after row=\bottomrule}}

\begin{table} [ht]
	\footnotesize
	\centering
	\pgfplotstabletypesetfile{Data/sweepd4DvsCdense.txt}
	\caption{Influence of $d$ for coupled (C) vs. decoupled (D) strategies in {\SpikeHyb} ($N= \SI{200000}{}$, $P=50$, $K=200$). All timings are in milliseconds. Symbols used are as specified for Table \ref{t:sweepP4DvsCdense}.} 
	\label{t:sweepd4DvsCdense}
\end{table}

\subsubsection{Comparison with Intel's MKL over a spectrum of $N$ and $K$}
\label{sss:sensitivityWRT-NK}
This section summarizes results of a two-dimensional sweep over $N$ and $K$. In this exercise, prompted by the results reported in Figs.~\ref{f:P-sweep}~and \ref{f:SPIKE_MKL_banded:d}, we fixed $P=50$ and chose matrices for which $d=1$. Each row in Table~\ref{t:denseBandedSweep} lists the value of $N$, which runs from \SI{1000}{} to \SI{1000000}{}. Each column lists the dimension of half bandwidth $K$, which runs from \SI{10}{} to \SI{500}{}. Each table row is split in three sub-rows: {\SaPD} results are reported in the first sub-row;  {\SaPC} in the second sub-row; {\MKL} in the third sub-row. All timings are in milliseconds. ``OOM'' stands for ``out-of-memory'' -- a situation that arises when {\SpikeHyb} exhausts during the solution of the linear system the GPU's 6 GB of global memory. 

\begin{table}[ht]
	\footnotesize
\centering
	\begin{tabular} {ccccccc}
		\hline
		\multirow{2}{*}{$N$} & \multicolumn{6}{c}{$K$} \rule{0pt}{2.6ex}\\ \cline{2-7}
		                     & 10 & 20 & 50 & 100 & 200 & 500 \rule{0pt}{2.6ex}\\ \hline
		\multirow{3}{*}{$\SI{1000}{}$}       & $\SciNum{24.33}$   & $\SciNum{17.546}$   & $\SciNum{18.163}$  & $\SciNum{20.665}$ & $\SciNum{27.552}$      & $\SciNum{29.515}$  \\ %\cline{2-7}
											 & $\SciNum{6.637}$   & $\SciNum{7.354}$    & $\SciNum{11.063}$  & $\SciNum{18.662}$ & $\SciNum{29.366}$      & $\SciNum{29.551}$  \\ %\cline{2-7}
											 & $\SciNum{11.4533}$ & $\SciNum{10.79596667}$   &    $\SciNum{12.80526667}$   &  $\SciNum{22.0766}$   &      $\SciNum{214.4676667}$   &      $\SciNum{220.817}$   \\ \hline
		\multirow{3}{*}{$\SI{2000}{}$}       & $\SciNum{22.242}$  & $\SciNum{18.731}$   & $\SciNum{19.109}$  & $\SciNum{21.488}$ & $\SciNum{27.253}$      & $\SciNum{56.376}$  \\ %\cline{2-7}
											 & $\SciNum{6.158}$   & $\SciNum{8.514}$    & $\SciNum{13.279}$  & $\SciNum{24.644}$ & $\SciNum{35.692}$      & $\SciNum{95.135}$  \\ %\cline{2-7}
											 & $\SciNum{12.54566667}$   &  $\SciNum{10.9954}$   &      $\SciNum{13.23996667}$   &  $\SciNum{22.14473333}$   &  $\SciNum{221.4436667}$   &  $\SciNum{235.6806667}$   \\ \hline
		\multirow{3}{*}{$\SI{5000}{}$}       & $\SciNum{25.17}$   & $\SciNum{20.622}$   & $\SciNum{21.012}$  & $\SciNum{23.272}$ & $\SciNum{32.592}$      & $\SciNum{80.022}$  \\ %\cline{2-7}
											 & $\SciNum{7.597}$   & $\SciNum{9.266}$    & $\SciNum{16.219}$  & $\SciNum{30.486}$ & $\SciNum{58.659}$      & $\SciNum{237.245}$  \\ %\cline{2-7}
											 & $\SciNum{13.07463333}$   &  $\SciNum{12.32803333}$   &  $\SciNum{21.44596667}$   &  $\SciNum{38.26973333}$   &  $\SciNum{253.058}$   &      $\SciNum{294.417}$   \\ \hline
		\multirow{3}{*}{$\SI{10000}{}$}      & $\SciNum{28.234}$  & $\SciNum{27.577}$   & $\SciNum{23.853}$  & $\SciNum{26.858}$ & $\SciNum{45.087}$      & $\SciNum{118.29}$  \\ %\cline{2-7}
											 & $\SciNum{10.192}$  & $\SciNum{11.678}$   & $\SciNum{18.873}$  & $\SciNum{45.607}$ & $\SciNum{105.997}$      & $\SciNum{473.732}$  \\ %\cline{2-7}
											 & $\SciNum{15.59736667}$      &  $\SciNum{15.08643333}$      &  $\SciNum{29.59363333}$      &  $\SciNum{58.80506667}$      &  $\SciNum{300.8843333}$      &  $\SciNum{392.8296667}$    \\ \hline
		\multirow{3}{*}{$\SI{20000}{}$}      & $\SciNum{33.926}$  & $\SciNum{32.35}$    & $\SciNum{33.019}$  & $\SciNum{41.984}$ & $\SciNum{59.909}$      & $\SciNum{201.6}$  \\ %\cline{2-7}
											 & $\SciNum{14.276}$  & $\SciNum{16.532}$   & $\SciNum{27.413}$  & $\SciNum{66.757}$ & $\SciNum{195.034}$      & $\SciNum{950.038}$  \\ %\cline{2-7}
											 & $\SciNum{20.86733333}$  &  $\SciNum{23.23023333}$  &     $\SciNum{48.79496667}$  &       $\SciNum{111.6733333}$  &      $\SciNum{337.3146667}$  &     $\SciNum{594.714}$  \\ \hline
		\multirow{3}{*}{$\SI{50000}{}$}      & $\SciNum{64.332}$  & $\SciNum{58.251}$   & $\SciNum{58.694}$  & $\SciNum{90.853}$ & $\SciNum{146.588}$     & $\SciNum{436.112}$  \\ %\cline{2-7}
											 & $\SciNum{27.125}$  & $\SciNum{30.484}$   & $\SciNum{54.7}$    & $\SciNum{144.387}$ & $\SciNum{366.771}$      & $\SciNum{2336.61}$  \\ %\cline{2-7}
											 & $\SciNum{32.63}$  &           $\SciNum{41.072}$  &          $\SciNum{102.97}$  &            $\SciNum{259.713}$  &           $\SciNum{715.1213333}$  &       $\SciNum{1106.736667}$  \\ \hline
		\multirow{3}{*}{$\SI{100000}{}$}     & $\SciNum{98.382}$  & $\SciNum{87.029}$   & $\SciNum{111.235}$ & $\SciNum{152.657}$ & $\SciNum{291.707}$    & $\SciNum{957.08}$  \\ %\cline{2-7}
											 & $\SciNum{47.645}$  & $\SciNum{55.755}$   & $\SciNum{96.496}$  & $\SciNum{261.166}$ & $\SciNum{649.768}$      & $\SciNum{3583.26}$  \\ %\cline{2-7}
											 & $\SciNum{53.921}$  &          $\SciNum{69.66366667}$  &     $\SciNum{191.046}$  &          $\SciNum{495.563}$  &          $\SciNum{1275}$  &             $\SciNum{2276.77}$  \\ \hline
		\multirow{3}{*}{$\SI{200000}{}$}     & $\SciNum{180.805}$ & $\SciNum{159.035}$  & $\SciNum{187.717}$ & $\SciNum{328.483}$ & $\SciNum{567.937}$    & $\SciNum{2002.77}$  \\ %\cline{2-7}
											 & $\SciNum{89.921}$  & $\SciNum{103.472}$  & $\SciNum{186.756}$ & $\SciNum{505.424}$ & $\SciNum{1221.29}$      & $\SciNum{6051.28}$  \\ %\cline{2-7}
											 & $\SciNum{95.08926667}$  &     $\SciNum{125.8573333}$  &    $\SciNum{367.6383333}$  &      $\SciNum{983.145}$  &         $\SciNum{2385.536667}$  &      $\SciNum{4210.51}$  \\ \hline
		\multirow{3}{*}{$\SI{500000}{}$}     & $\SciNum{371.996}$ & $\SciNum{365.062}$  & $\SciNum{442.454}$ & $\SciNum{724.046}$ & $\SciNum{1410.76}$    &     OOM       \\ %\cline{2-7}
											 & $\SciNum{203.744}$ & $\SciNum{237.991}$  & $\SciNum{442.407}$  & $\SciNum{1229.47}$ & $\SciNum{2928.33}$      &       OOM     \\ %\cline{2-7}
											 & $\SciNum{213.471}$ &        $\SciNum{292.442}$ &        $\SciNum{896.9266667}$ &      $\SciNum{2539.273333}$ &      $\SciNum{6231.033333}$ &      $\SciNum{10706.63333}$ \\ \hline
		\multirow{3}{*}{$\SI{1000000}{}$}    & $\SciNum{724.221}$ & $\SciNum{709.176}$  & $\SciNum{978.835}$ & $\SciNum{1442.33}$ &            OOM           &       OOM     \\ %\cline{2-7}
											 & $\SciNum{396.964}$ & $\SciNum{463.345}$  & $\SciNum{863.955}$ & $\SciNum{2442.78}$ &          OOM             &       OOM       \\ %\cline{2-7}
											 & $\SciNum{348.634}$ &        $\SciNum{569.2476667}$ &    $\SciNum{1777.773333}$ &      $\SciNum{4712.44}$ &          $\SciNum{11367.23333}$ &      $\SciNum{21592.06667}$  \\ \hline
	\end{tabular}
	\caption{Performance comparison, two-dimensional sweep over $N$ and $K$ for $P=50$ and $d=1$. For each value $N$, the three rows correspond to the {\SaPD}, {\SaPC}, and {\MKL} solvers, respectively.}
\label{t:denseBandedSweep}
\end{table}

\begin{figure}
	\centering
	\includegraphics[width=0.6\textwidth]{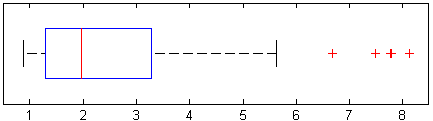}
	\caption{{\SaP} speedup over Intel's MKL -- statistical analysis based on values in Table \ref{t:denseBandedSweep}.}
	\label{fig:denseMatSpdUpMKL}
\end{figure}

The results reported in Table~\ref{t:denseBandedSweep} are statistically summarized in Fig.~\ref{fig:denseMatSpdUpMKL}, which provides {\SaP} over {\MKL} speedup information. Assume that a test ``${\alpha}$'' successfully ran to completion in {\SaPD}, requiring $T^{\SaPD}_{\alpha}$, and/or in {\SaPC}, requiring $T^{\SaPC}_{\alpha}$. By convention, in case of failing to solve, a negative value; i.e. -1, is assigned to $T^{\SaPD}_{\alpha}$ or $T^{\SaPC}_{\alpha}$. If a test runs to completion both in {\SaP} and {\MKL}, the ``${\alpha}$'' speedup value used to generate the plot in Fig.~\ref{fig:denseMatSpdUpMKL} is computed as $s_{BD} \equiv T^{\MKL}_{\alpha}/T^{\SaP}_{\alpha}$, where $T^{\MKL}_{\alpha}$ is {\MKL}'s time to solution and $T^{\SaP}_{\alpha} \equiv \min(\max(T^{\SaPD}_{\alpha},0),\max(T^{\SaPC}_{\alpha},0))$. 
Given that $N$ assumes 10 values and $K$ takes 6 values, ``${\alpha}$'' can be one of 60 tests. Since three $(N,K)$ tests, namely $(\SI{1000000}{},200)$, $(\SI{1000000}{},500)$, and $(\SI{500000}{},500)$, failed to solve in {\SaP}, the sample population for the statistical study in Fig.~\ref{fig:denseMatSpdUpMKL} is 57. Out of 57 tests, $s_{BD}>1$ in all but two cases: for $(\SI{1000000}{},10)$ when $s_{BD}=0.87825$, and for $(\SI{2000}{},50)$ when $s_{BD}=0.99706$. The highest speedup was $s_{BD}=8.1255$, for $(\SI{2000}{},200)$. The median is slightly higher than 2.0, which indicates that of the 57 tests, half were completed by {\SaP} two times faster than by {\MKL}. The figure also shows that about 25\% of the tests run, roughly, between three and six times faster in {\SaP}. The red crosses in the figure represent outliers.

%\FloatBarrier
%-------------------------------------------------------------------------------
\subsection{Numerical experiments related to sparse matrix reorderings}
\label{ss:reorderingNumExp}
When solving sparse linear systems, {\SaP} reformulates the sparse problem as a dense banded linear system that is subsequently solved using {\SaPC} or {\SaPD}. Ideally, the ``sparse--to--dense'' transition yields a coefficient matrix that is diagonal heavy; i.e., has a large $d$, and has a small bandwidth $K$. Two matrix reorderings are applied in an attempt to meet these two objectives. The first one; i.e., the diagonal boosting reordering, is assessed in section \S\ref{sss:diagBoostReord}. The second one; i.e., the bandwidth reduction reordering, is evaluated in \S\ref{sss:BWreduction}.

\subsubsection{Assessment of the diagonal boosting reordering solution}
\label{sss:diagBoostReord}
\begin{figure}
\centering
\includegraphics[width=\textwidth]{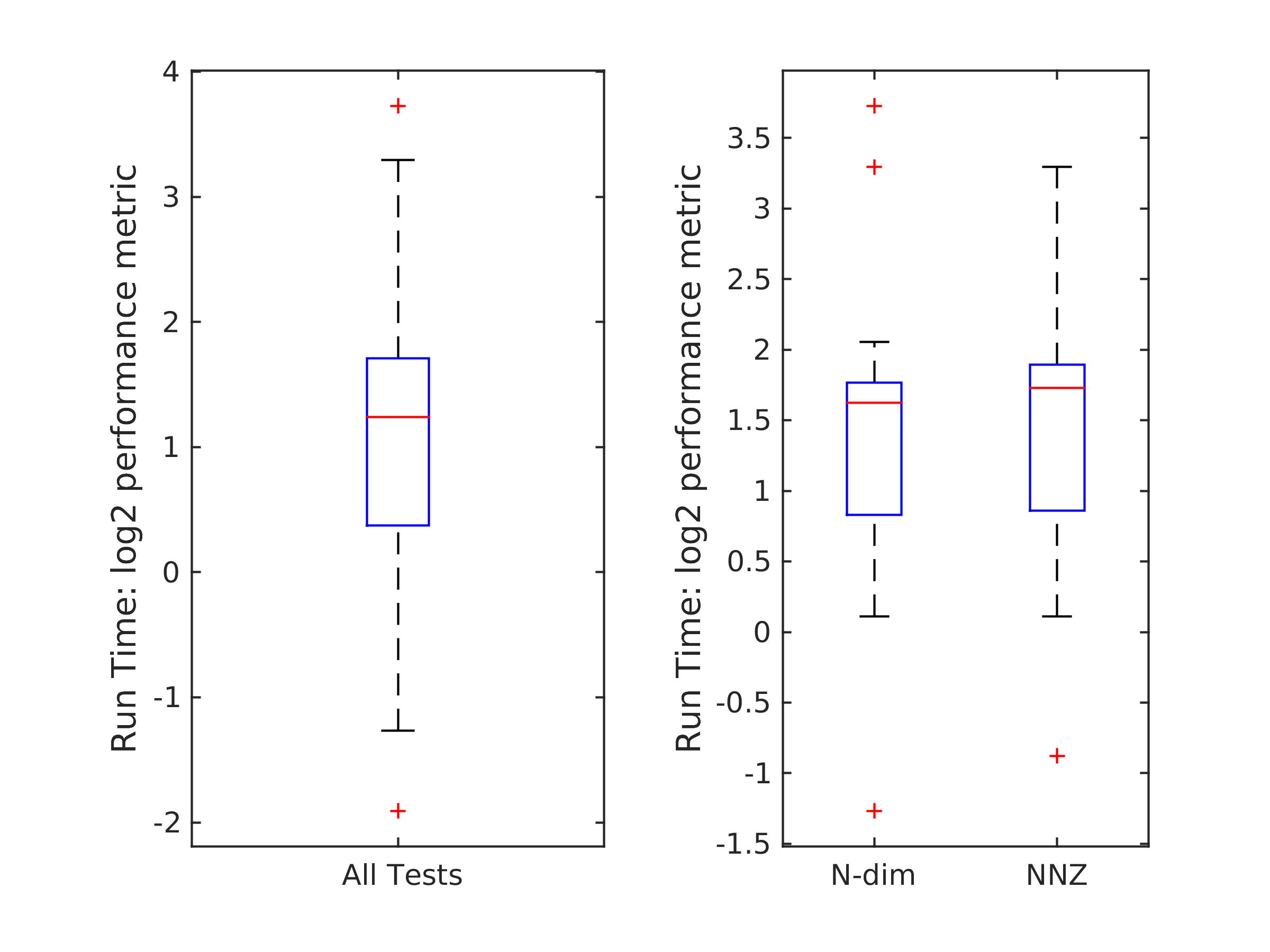}
\caption{Results of a statistical analysis that uses a median-quartile method to measure the spread of the {\MCsf} and {\DB} times to solution. The speedup factor, or performance metric, is computed as in Eq.~(\ref{eq:spreadDefinitionDB}).}
\label{fig:db-speedup}
\end{figure}

The first set of results, summarized in Fig. \ref{fig:db-speedup}, correspond to an efficiency comparison between the hybrid CPU--GPU implementation of \S\ref{ss:db-impl} and the Harwell Sparse Library (HSL) {\MCsf} algorithm \cite{HSL}. The hybrid implementation outperformed {\MCsf} for 96 out of the 116 matrices selected from the Florida Sparse Matrix Collection~\cite{davis2011university}. The left pane in Fig.~\ref{fig:db-speedup} presents results of a statistical analysis that used a median-quartile method to measure the spread of the {\MCsf} and {\DB} times to solution. Assume that $T_\alpha^{\DB}$ and $T_\alpha^{\MCsf}$ represent the times required by {\DB} and {\MCsf}, respectively, to complete the diagonal boosting reordering in test $\alpha$. A relative speedup is computed as
\begin{equation}
{\cal S}_\alpha^{\DB-\MCsf} = 
\log_2 \frac{T_\alpha^{\MCsf}}{T_\alpha^{\DB}} \, .
%\left\{ {
%	\begin{aligned}
%	\log_2 \frac{T_\alpha^{\MCsf}}{T_\alpha^{\DB}} &\quad \text{if} \quad  T_\alpha^{\MCsf}>T_\alpha^{\DB} \vspace{0.25cm}\\
%	-\log_2 \frac{T_\alpha^{\DB}}{T_\alpha^{\MCsf}} &\quad \text{if} \quad T_\alpha^{\DB} \geq T_\alpha^{\MCsf}
%	\end{aligned}
%}\right.
%\quad .
\label{eq:spreadDefinitionDB}
\end{equation}
\noindent These ${\cal S}_\alpha^{\DB-\MCsf}$ values, which can be either positive or negative, are collected in a set ${\cal S}^{\DB-\MCsf}$ which is used to generate the left box plot in Fig.~\ref{f:boxPlotSparseSolvers}. The number of tests used to produce these statistical results was 116. Note that a positive value means that {\DB} is faster than {\MCsf}, with the opposite outcome being the case for negative values of ${\cal S}_\alpha^{\DB-\MCsf}$. The median values for ${\cal S}^{\DB-\MCsf}$ was $1.2423$, which indicates that half of the 116 tests ran more than 2.3 times faster using the {\DB} implementation. On average, it turns out that the larger the matrix, the faster the {\DB} solution becomes. Indeed, as a case study, we analyzed a subset of larger matrices. The ``large'' attribute was defined in two ways: first, by considering the matrix size, and second, by considering the number of nonzero elements. For the 116 matrices considered, we picked the largest 24 of them; i.e., approximately the largest 20\%. To this end, in the first case, we selected all matrices whose dimension was higher than $N=$\SI{150000}{}. In the second case, we selected all matrices whose number of nonzero elements was larger than \SI{4350000}{}. For large $N$, the median was 1.6255, while for matrices with many nonzero elements, the median was 1.7276. In other words, half of the large tests ran more than three times faster in {\DB}. Finally, the statistical results in Fig.~\ref{f:boxPlotSparseSolvers} indicate that for large tests, with the exception of two outliers, there were no tests for which ${\cal S}_\alpha^{\DB-\MCsf}$ was negative; i.e., with one exception, {\DB} was faster. When all 116 tests were considered, {\MCsf} was faster in several cases, with an outlier for which {\MCsf} was four times faster than {\DB}. 

Two facts emerged at the end of this analysis. First, as discussed in \cite{AngMC64-2014}, the bottleneck in the diagonal boosting reordering was either the {\DBS{2}} stage; i.e., finding the initial match, or the {\DBS{3}} stage; i.e., finding a perfect match, with an approximately equal split among them. Secondly, the quality of the reordering turned out to be identical -- almost all matrices displayed the same grand product of the diagonal entries regardless of whether the reordering was carried out using {\MCsf} or {\DB}.

\subsubsection{Assessment of the bandwidth reduction solution}
\label{sss:BWreduction}

\begin{figure} [ht]
	\centering
	\includegraphics[width=0.8\textwidth]{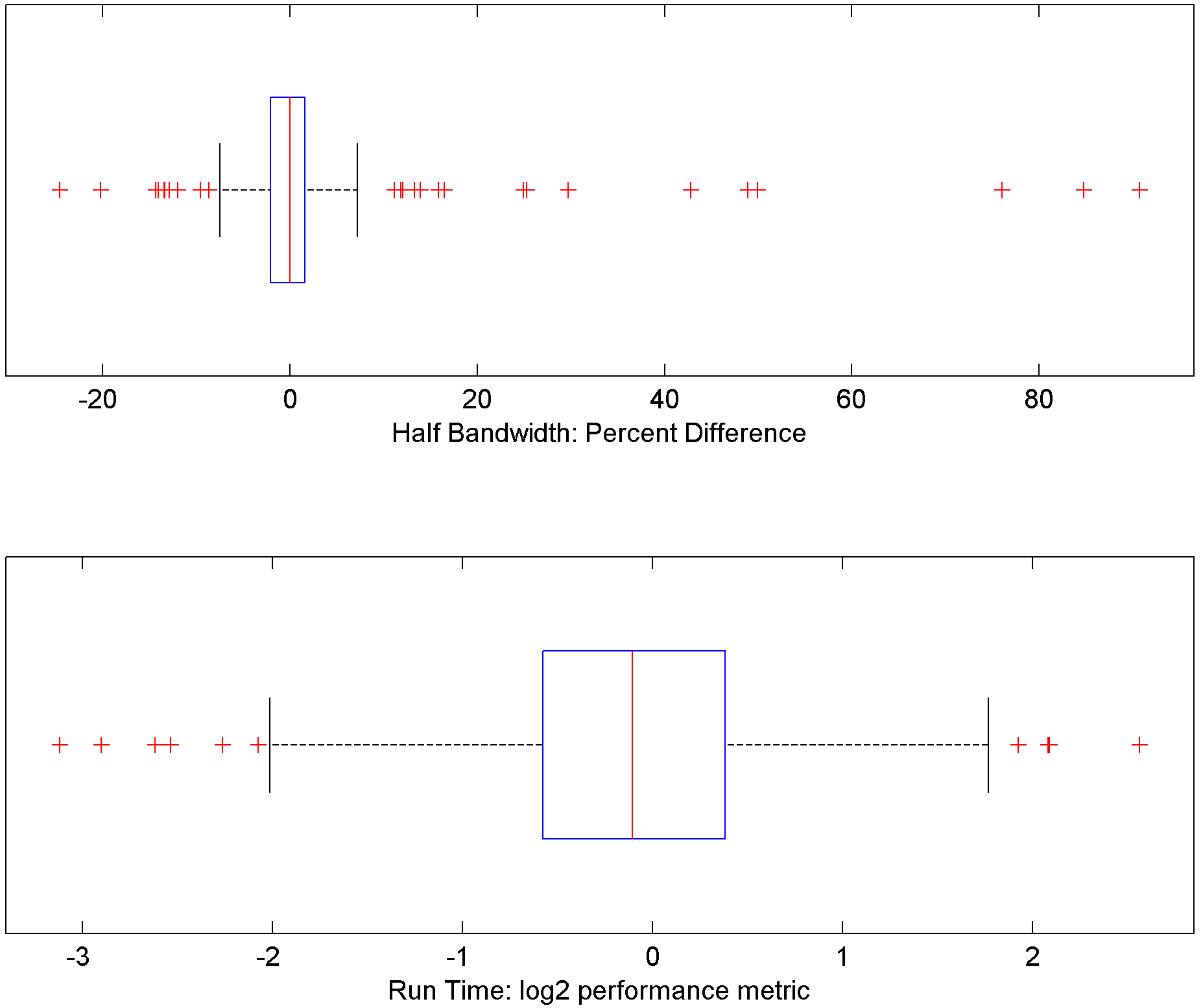}
	\caption{Comparison of the Harwell {\MCs} and {\SaP}'s {\CM} implementations in terms of resulting half bandwidth $K$ and time to solution.}
	\label{fig:allMatResCM}
\end{figure}

The performance of the {\CM} solution implemented in {\SaP} was evaluated on a set of 125 sparse matrices from various applications. These matrices were the 116 used in the previous section plus several other matrices such as ANCF31770, ANCF88950, and NetANCF\_40by40, etc., that arise in granular dynamics and the implicit integration of flexible multi-body dynamics~\cite{luningThesis2015,LuningTechReport2014,serban2015}. Figure~\ref{fig:allMatResCM} presents results of a statistical analysis that used a median-quartile method to compare ($i$) the half bandwidths of the matrices obtained by Harwell's {\MCs} and {\SaP}'s {\CM}; and, ($ii$) the time to solution; i.e., time to complete a band-reducing reordering. For ($i$), the quantity reported is the relative difference between the resulting bandwidths,
\[
r_K \equiv 100 \times \frac{K_{\MCs}-K_{\CM}}{K_{\CM}} \, ,
\]
where $K_{\MCs}$ and $K_{\CM}$ are, respectively, the half bandwidths $K$ of the matrices produced by {\MCs} and {\CM}. For ($ii$), the metric used was identical to the one introduced in Eq.~(\ref{eq:spreadDefinitionDB}). Note that {\CM} is superior when $r_K$ assumes large positive values, which are also desirable for the time-to-solution plot. As far as $r_K$ is concerned, the median value is $0\%$; i.e., out of 125 matrices, about half are better off being reordered by Harwell's {\MCs} with the other half being better off reordered by {\SaP}'s {\CM}. On a positive side, the number of outliers for {\CM} is higher, indicating that there is a propensity for {\CM} to ``win big''. In terms of times to solution, {\MCs} is marginally faster than {\CM}'s hybrid CPU/GPU solution. Indeed, the median value of the performance metric is $-0.1057$; i.e., it takes half of the tests run with {\CM} at least $1.076$ times longer to complete the bandwidth reduction task. 

\begin{figure} [ht]
\centering
\includegraphics[width=0.8\textwidth]{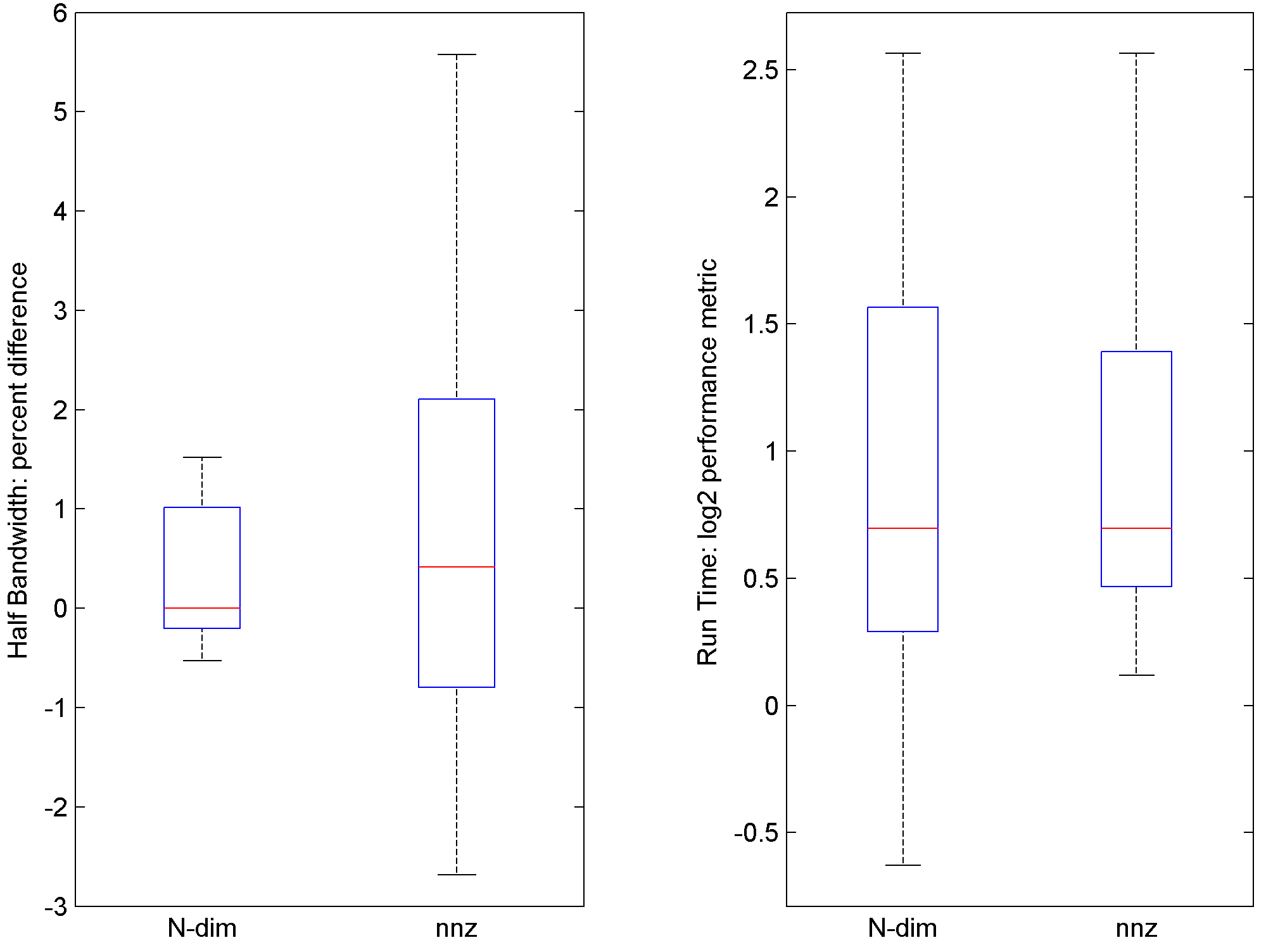}
\caption{Comparison of the Harwell {\MCs} and {\SaP}'s {\CM} implementations in terms of resulting half bandwidth $K$ and time to solution. Statistical analysis of large matrices only.}
\label{fig:largeMatResCM}
\end{figure}

It is insightful to discuss what happens when this statistical analysis is controlled to only consider larger matrices. The results of this analysis are captured in Fig.~\ref{fig:largeMatResCM}. Just like in section \S\ref{sss:diagBoostReord}, the focus is on the largest 20\% matrices, where ``large'' is understood to mean large matrix dimension $N$, and then separately, large number of nonzeros $nnz$. Incidentally, the cut-off value for the dimension was $N=$\SI{215000}{}, while for the number of nonzeros was $nnz=$\SI{7800000}{}. When the statistical analysis included the 25 largest matrices based on size $N$, the median value for the half bandwidth metric $r_K$ was yet again $0.0\%$. The median value for time to solution changed however, from $-0.1057$ to $0.6964$ to indicate that for half of these large tests {\SaP} ran more than $1.6$ times faster than the Harwell solution. Qualitatively, the same conclusions were reached when the 25 large matrices were selected on the grounds on $nnz$ count. The median for $r_K$ was $0.4182\%$, which again suggested that the relative difference in the resulting bandwidth $K$ yielded by {\CM} and {\MCs} was practically negligible. The median time to solution was the same $0.6964$. Note though that according to the results shown in Fig.~\ref{fig:largeMatResCM}, there is no large--$nnz$ test for which the Harwell implementation is faster than the {\CM}. In fact, 25\% of the large tests; i.e., about five tests, run at least three times faster in {\CM}.

Finally, it is worth pointing out the correlations between times to solutions and $K$ values, on the one hand, and $N$ and $nnz$, on the other hand. Herein, the correlation used is the Pearson product-moment correlation coefficient \cite{Box1978}. As a rule of thumb, a Pearson correlation coefficient of 0.01 to 0.19 suggests a negligible relationship, while a coefficient between 0.7 and 1.0 indicates a strong positive relationship. The correlation coefficient between the bandwidth and the dimension $N$ of the matrix turns out to be small; i.e., $0.15$ for {\MCs} and $0.16$ for {\CM}. Indeed, the fact that a matrix is large doesn't say much about what $K$ value one can expect upon reordering this matrix. The correlation between the number of nonzeros and the amount of time to figure out the reordering is very high though. In other words, the larger the matrix size $N$, the longer the time to produce the reordering. For instance, the correlation coefficient was $0.91$ for {\MCs} and $0.81$ for {\CM}. The same observation holds for the number of nonzeros entries: when there is a lot of them, the time to produce a reordering is large. The Pearson correlation coefficient is $0.71$ for {\MCs} and $0.83$ for {\CM}. These correlation coefficients were obtained on a sample size of 125 matrices. Yet the same trends are manifest for the reduced set of 25 large matrices that we worked with. For instance, the correlation between dimension $N$ and resulting $K$ is very small at large $N$ values: $0.04$ for {\MCs} and $0.05$ for {\CM}. For the time to solution, the correlation coefficients with respect to $N$ are $0.89$ for {\MCs} and $0.76$ for {\CM}.

\FloatBarrier

\subsection{Numerical experiments related to sparse linear systems}
\label{ss:sparseLinSysExp}

\subsubsection{Profiling results}
\label{sss:profilingStudy}
\begin{figure}[ht]
	\centering {\includegraphics[width=0.85\textwidth]{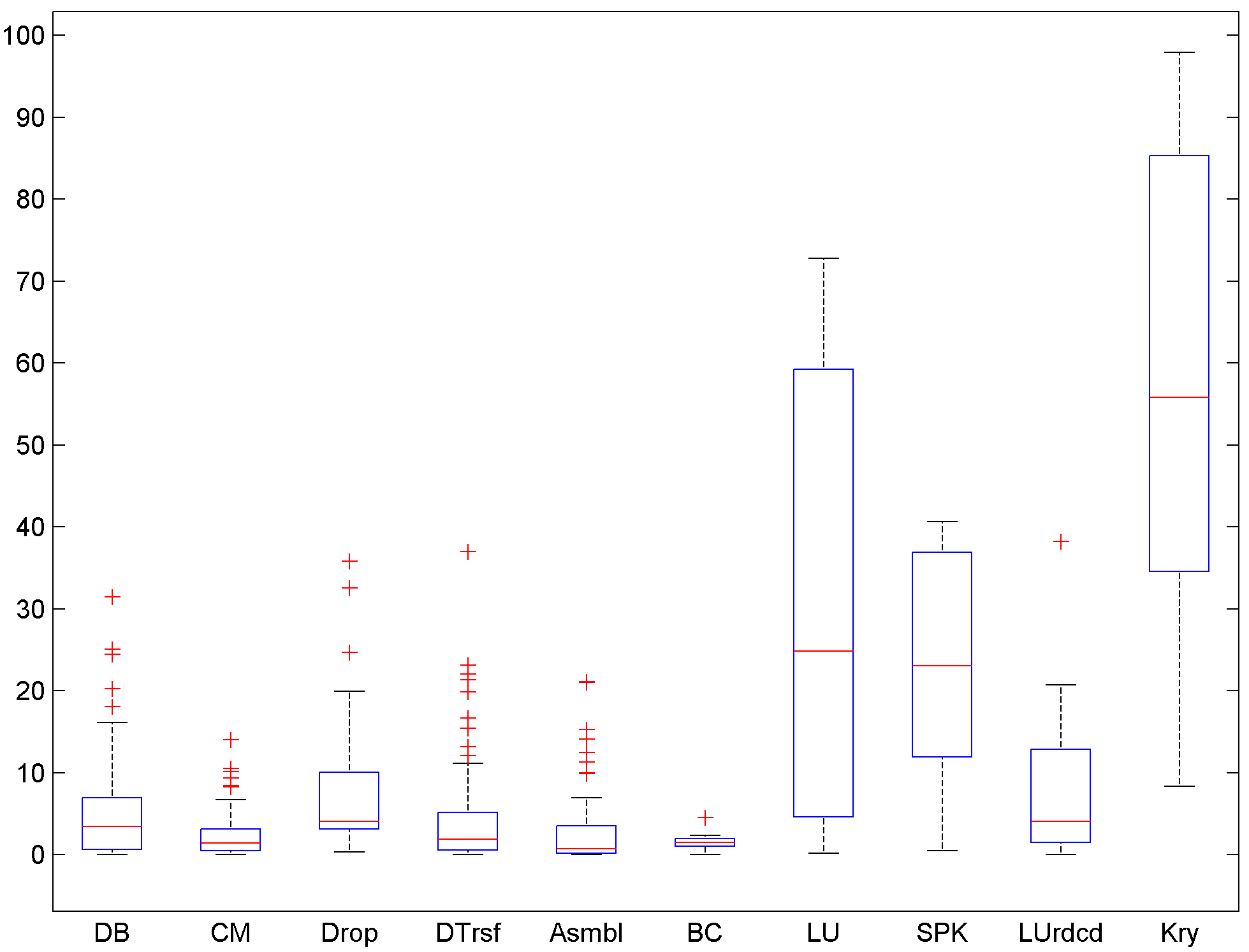}}
	\caption{Profiling results obtained for a set of 85 linear systems that, out of a collection of 114, could be solved by {\SpikeHyb}.}
	\label{f:profilingResults}
\end{figure}

Figure~\ref{f:profilingResults} plots statistical results that summarize how the time to solution; i.e., finding $\bx$ in $\bA \bx = \bb$, is spent in {\SpikeHyb}. The raw data used in this analysis is available on-line  \cite{SaPWebsite}; also, a discussion of exactly what it means to find the solution of the linear system is postponed for section \S\ref{sss:compareAgainst-cuSolver}. The labels used in the plot Fig.~\ref{f:profilingResults} are inspired by the notation used in section \S\ref{ss:overallAlg} and Fig.~\ref{f:comFlowSpike}. Consider for instance the diagonal boosting reordering {\DB} employed by {\SaP}. In a statistical sense, the percent of time to solution spent in {\DB} is represented using a median-quartile method to measure statistical spread. The raw data used to generate the {\DB} box was obtained as follows. If a test ``${\alpha}$'' that runs to completion requires $T^{DB}_{\alpha}>0$ for {\DB} completion, then this test will generate one data entry in an array of data subsequently used to produce the statistical result. The actual entry that is used is $100 \times T^{DB}_{\alpha}/T^{Tot}_{\alpha}$, where $T^{Tot}_{\alpha}$ is the total amount of time that test ``${\alpha}$'' takes for completion. In other words, the entry is the percent of time spent when solving this particular linear system for performing the diagonal boosting reordering. The bars for the $K$-reducing reordering ({\CM}), for multiple data transfers between CPU and GPU ({\tt Dtrsf}), etc., are similarly obtained. 
Not all bars in Fig.~\ref{f:profilingResults} were generated using the same number of data entries; i.e., some tests contributed to some but not all bars. For instance, a symmetric positive definite linear system requires no {\DB} step and such this test won't contribute an entry to the array of data used to determine the {\DB} box in the figure. Of a batch of 85 tests that ran to completion with {\SaP}, the sample population used to generate the bars is as follows: 85 data points for {\CM}, {\tt Dtrsf}, and {\tt Kry}; 63 data points for {\DB}; 60 for {\tt LU}; 32 data points for {\tt Drop}; and 9 data points for {\tt BC}, {\tt SPK}, and {\tt LUrdcd}. These counts provide insights into the solution path adopted by {\SaP} in solving the 85 linear systems. For instance, the coupled approach; i.e., the SPIKE method of \cite{PoSa2006} has been employed in the solution of nine of the 85 linear systems. The rest of them were used via {\SaPD}. Of 85 linear systems, 25 were most effectively solved by {\SaP} resorting to diagonal preconditioning; i.e., after {\DB} all the entries were dropped off except the heavy diagonal ones. Also, note that several of the linear systems considered were symmetric positive definite, from where the 60 points count for {\DB}. 

\begin{figure}[ht]
	\centering {\includegraphics[width=0.85\textwidth]{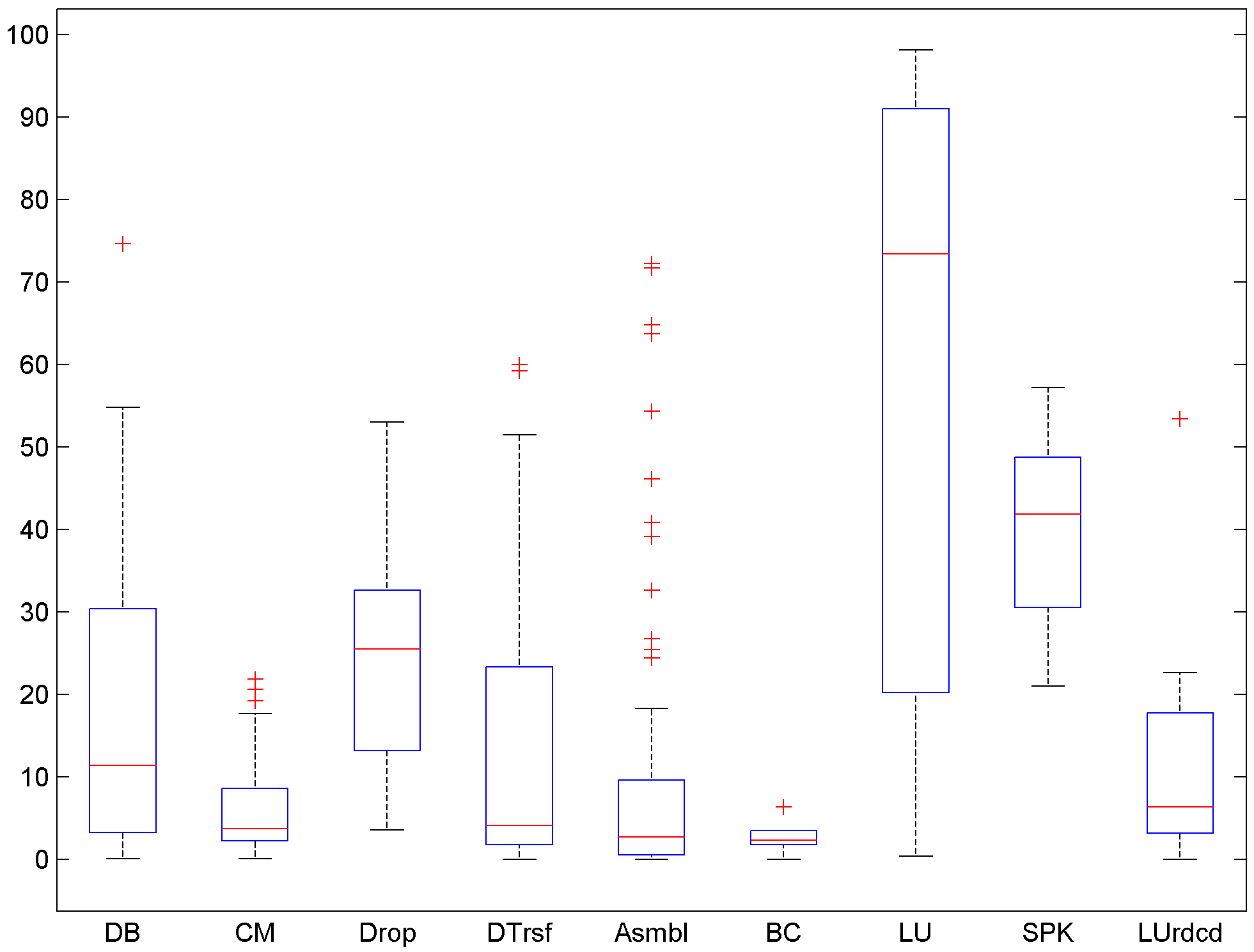}}
	\caption{Profiling results obtained for a set of 85 linear systems that, out of a collection of 114, could be solved by {\SpikeHyb}.}
	\label{f:profileSaP_noKryl}
\end{figure}

A statistical analysis of the time spent in the Krylov-subspace component of the solution reveals that the median time was 55.84\%. The median times for the other components of the solution are listed in the first row of data in  Table~\ref{t:medianSaPcomponents}. The second row of data provides the median values when the Krylov-subspace component, which dwarfs most of the solution components is eliminated. In this case, the entry for {\DB}, for instance, was obtained based on data points $100 \times T^{DB}_{\alpha}/T^{Tot}_{\alpha}$, where this time around $T^{Tot}_{\alpha}$ included everything except the time spent in the Krylov-subspace component of the solution. In other words, $T^{Tot}_{\alpha}$ is the time required to compute from scratch the preconditioner. The median values should be used in conjunction with the median-quartile boxplot of Fig.~\ref{f:profilingResults} for the first row of data, and Fig.~\ref{f:profileSaP_noKryl} for the second row of data. Consider, for instance, the results associated with the drop-off operation. In the Krylov-inclusive measurement, {\tt Drop} has a median of 4.1\%; i.e., half of the 32 tests which employed drop-off spent more than amount in performing the drop-off, while half were quicker. The spread is rather large and there are several outliers that suggest that a handful of tests require a very large amount of time be spent in the drop-off part of the solution.

\pgfplotstableset{
	columns={DB,CM,Dtransf,Drop,Asmbl,BC,LU,SPK,LUrdcd},
	columns/DB/.style={
		fixed,
		column name={\DB},
		precision=1},
	columns/CM/.style={
		column name={\CM},
		fixed,
		precision=1},
	columns/Dtransf/.style={
		column name={\tt Dtransf},
		fixed,
		precision=1},
	columns/Drop/.style={
		column name={\tt Drop},
		fixed,
		precision=1},
	columns/Asmbl/.style={
		column name={\tt Asmbl},
		fixed,
		precision=1},
	columns/BC/.style={
		column name={\tt BC},
		fixed,
		precision=1},
	columns/LU/.style={
		column name={\tt LU},
		fixed,
		precision=1},
	columns/SPK/.style={
		column name={\tt SPK},
		fixed,
		precision=1},
	columns/LUrdcd/.style={
		column name={\tt LUrdcd},
		fixed,
		precision=1},
%	columns/Kry/.style={
%		column name={\tt Kry},
%		fixed,
%		precision=1},
	every head row/.style={
		before row=\toprule,after row=\midrule},
	every last row/.style={
		after row=\bottomrule}}

\begin{table} [ht]
	\footnotesize
	\centering
	\pgfplotstabletypesetfile{Data/medianSaPcomponents.txt}
	\caption{Median information for the {\SaP} solution components as \% of the time for solution. Two scenarios are considered: the first data row provides values when the total time; i.e., 100\%, included the time spent by {\SaP} in the Krylov-subspace component. The second row of data is obtained by considering 100\% to be the time required to compute a factorization of the preconditioner. Note that values in each row of data does not add up to 100\% for several reasons. First, these are statistical median values. Second, there are very many tests that do not include all the components of the solution. For instance, {\tt SPK} is computed based on a set of nine points while {\tt Drop} is computed using 32 data points, some of them not even obtained in conjunction with the same test.} 
	\label{t:medianSaPcomponents}
\end{table}

% -----------------
% -----------------

% -----------------
% -----------------

The results in Fig.~\ref{f:profilingResults} and Table~\ref{t:medianSaPcomponents} suggest where the optimization efforts should concentrate in the future. For instance, the time required for the CPU$\leftrightarrow$GPU data transfer is, in the overall picture, rather insignificant and as such a matter of small concern. 
Somewhat surprising, the amount of time spent in drop-off came out higher than anticipated, at least in relative terms. One caveat is that no effort was made to optimize this component of the solution. Instead, the effort went into optimizing the {\DB} and {\CM} solution components. This paid off, as matrix reordering in {\SaP}, particularly for large matrices, is fast when compared to Harwell and it reached the point where the drop-off became a more significant bottleneck. Another unexpected observation was the relative small number of scenarios in which {\SaPC} was preferred over {\SaPD}; i.e., in which the SPIKE strategy \cite{PoSa2006}  was employed. This observation, however, should not be generalized as it might very well be specific to the {\SaP} implementation. Indeed, it simply states that in the current implementation, a large number of iterations associated with a less sophisticated preconditioner is preferred to a smaller count of expensive iterations associated with {\SaPC}. Out of a sample population of 85 tests, when invoked, the median number of iterations to solution in {\SaPC} was 6.75. Conversely, when {\SaPD} was preferred, the median count was 29.375 \cite{SaPWebsite}.

\subsubsection{The impact of the third stage reordering}
\label{sss:TSR}
It is almost always the case that upon carrying out a CM reordering of a sparse matrix, the resulting $\bA$ matrix has a small number of entries in the first and last rows. Yet, as the row index $j$ increases, the number of nonzero in row $j$ increases up to approximately $j \approx N/2$. Thereafter, the nonzero count starts decreasing to reach small values towards $j \approx N$. Overall, $\bA$ has its $K$ value dictated by the worst offender. Therefore, a partitioning of $\bA$ into $\bA_i$, $i=1,\ldots,P$ would conservatively require that, for instance, $\bA_1$ and $\bA_P$ work with a large $K$ most likely dictated by a sub-matrix such as $\bA_{P/2}$. Allowing each $\bA_i$ to have its own $K_i$ proved to lead to efficiency gains for two main reasons. First, in {\SaPC} it led to a reduction in the dimension of the spikes, since for each pair of coupling blocks ${\bB}_i$ and ${\bC}_i$, the number of columns in the ensuing spikes was determined as the larger of the values $K_i$ and $K_{i+1}$. Second, {\SpikeHyb} capitalizes on the observation that, since $\bA_i$ are independent and governed by their local $K_i$, there is nothing to prevent a third reordering, which attempts to further reduce the bandwidth of $\bA_i$. As it comes on the heels of the {\DB} and {\CM} reorderings, this is called a ``third stage reordering'' and is applied independently and preferably concurrently to the $P$ sub-matrices $\bA_i$. As illustrated in Table \ref{t:KsPs-3rdSR}, the decrease in local $K_i$ can be significant and it can lead to non-negligible speedups, see Table \ref{3rdSR-actualSpdUps}.

\begin{table}[ht]
	\footnotesize
	\centering
	\begin{tabular} {cccc}
	\hline
	Mat. Name                   &     $P$      & $K_i$ before {\TSR}  &   $K_i$ after {\TSR}    \rule{0pt}{2.6ex}  \\  \hline
	\multirow{4}{*}{ANCF31770}  & \multirow{4}{*}{$20$} & 123, 170, 204, 229, 247  &  89, 92, 79, 46, 45   \\ 
								&                       & 247, 247, 247, 248, 242  &  48, 48, 59, 50, 58   \\ 
	                            &                       & 213, 181, 134, 68, 106   &  72, 98, 64, 56, 42   \\ 
	                            &                       & 129, 124, 124, 113, 82   &  36, 54, 49, 59, 82   \\ \hline
	\multirow{4}{*}{ANCF88950}  & \multirow{4}{*}{$20$} & 194, 274, 337, 387, 410  & 116, 74, 65, 109, 112 \\
	                            &                       & 410, 410, 410, 410, 405  & 97, 100, 93, 97, 114  \\
	                            &                       & 352, 296, 227, 116, 176  & 116, 56, 88, 75, 116  \\
								&                       & 208, 204, 204, 191, 137  & 50, 96, 97, 118, 75   \\ \hline
	\multirow{2}{*}{af23560}    & \multirow{2}{*}{$10$} & 274, 317, 317, 317, 320  & 140, 71, 71, 102, 74    \\
	                            &                       & 339, 334, 317, 314, 283  & 123, 127, 119, 114, 143 \\ \hline
	\multirow{4}{*}{NetANCF40by40} & \multirow{4}{*}{$16$} & 256, 378, 458, 533    & 125, 68, 122, 118       \\
	                               &                       & 599, 634, 578, 517    & 85, 93, 97, 91          \\
	                               &                       & 436, 343, 215, 210    & 57, 69, 112, 85         \\
	                               &                       & 275, 295, 257, 178    & 85, 73, 113, 101        \\ \hline
	\multirow{2}{*}{bayer01}       & \multirow{2}{*}{$8$}  & 684, 1325, 1308, 1288 & 532, 170, 122, 110      \\
	                               &                       & 879, 501, 493, 508    & 109, 110, 110, 121      \\ \hline
	\multirow{2}{*}{ex19}          & \multirow{2}{*}{$8$}  & 139, 87, 87, 87       & 136, 87, 87, 87          \\
								   &                       & 74, 46, 62, 40        & 68, 46, 62, 40          \\ \hline
	\multirow{4}{*}{finan512}      & \multirow{4}{*}{$16$} & 1124, 1287, 1316, 1331 & 587, 288, 288, 288     \\
	                               &                       & 1331, 1331, 1331, 1331 & 288, 288, 288, 288     \\
	                               &                       & 1331, 1331, 1331, 1331 & 288, 288, 288, 288     \\
	                               &                       & 1331, 1331, 1331, 1015 & 288, 288, 227, 211     \\ \hline
	\multirow{2}{*}{gridgena}      & \multirow{2}{*}{$6$}  & 247, 405, 405          & 132, 81, 80            \\
								   &                       & 405, 405, 247          & 122, 72, 105           \\ \hline
	\multirow{2}{*}{lhr10c}        & \multirow{2}{*}{$6$}  & 315, 348, 288          & 427, 247, 293          \\
	                               &                       & 166, 156, 259          & 217, 226, 157          \\ \hline
	\multirow{2}{*}{rma10}         & \multirow{2}{*}{$10$} & 180, 281, 702, 678, 495 & 155, 241, 647, 540, 254 \\
	                               &                       & 637, 560, 495, 478, 545 & 496, 422, 217, 349, 358 \\ \hline
	\end{tabular}
	\caption{Examples of matrices where the third stage reordering ({\TSR}) reduced more significantly the block bandwidth $K_i$ for $\bA_i$, $i=1,\ldots,P$.}
	\label{t:KsPs-3rdSR}
\end{table}

\begin{table}[ht]
	\footnotesize
	\centering
	\begin{tabular} {cccccc}
		\hline
		\multirow{2}{*}{Mat. Name}     &    \multicolumn{2}{c}{w/o {\TSR}}   & \multicolumn{2}{c}{w/ {\TSR}}      & \multirow{2}{*}{SpdUp}   \\ \cline{2-5}
									   &     $P$                 & $K_i$                           &  $P$  & $K_i$  & \\ \hline
		ANCF31770     &    $16$         & $\SI{248}{}$  & $20$      & ${\SI{98}{}}$  & $1.203$ \\ \hline
		ANCF88950     &    $32$         & $\SI{410}{}$  & $20$      & ${\SI{118}{}}$ & $1.537$ \\ \hline
		af23560       &    $10$         & $\SI{339}{}$  & $10$      & ${\SI{143}{}}$ & $1.238$ \\ \hline
		NetANCF40by40 &    $16$         & $\SI{634}{}$  & $16$      & ${\SI{125}{}}$ & $1.900$ \\ \hline
		bayer01       &    $8$          & $\SI{1325}{}$ & $8$       & ${\SI{532}{}}$ & $2.234$ \\ \hline
		ex19          &    $6$          & $\SI{139}{}$  & $8$       & ${\SI{136}{}}$ & $1.331$ \\ \hline
		finan512      &    $10$         & $\SI{1331}{}$ & $16$      & $\SI{587}{}$   & $1.804$ \\ \hline
		gridgena      &    $6$          & $\SI{405}{}$  & $6$       & $\SI{132}{}$   & $1.636$ \\ \hline
		lhr10c        &    $4$          & $\SI{427}{}$  & $6$       & $\SI{259}{}$   & $1.228$ \\ \hline
		rma10         &    $10$         & $\SI{702}{}$  & $10$      & $\SI{647}{}$   & $1.113$ \\ \hline
	\end{tabular}
	\caption{Speed-up ``SpdUp'' values obtained upon embedding a third stage reordering step in the solution process, a decision that also changed the number of partitions $P$ for best performance. When correlating the results reported to values provided in Table \ref{t:KsPs-3rdSR}, this table lists for each matrix $\bA$ the largest of its $K_i$ values, $i=1,\ldots,P$.}
	\label{3rdSR-actualSpdUps}
\end{table}

\subsubsection{Comparison against state of the art}
\label{sss:compareAgainstOtherSS}
\begin{figure}[ht]
	\centering {\includegraphics[width=0.9\textwidth]{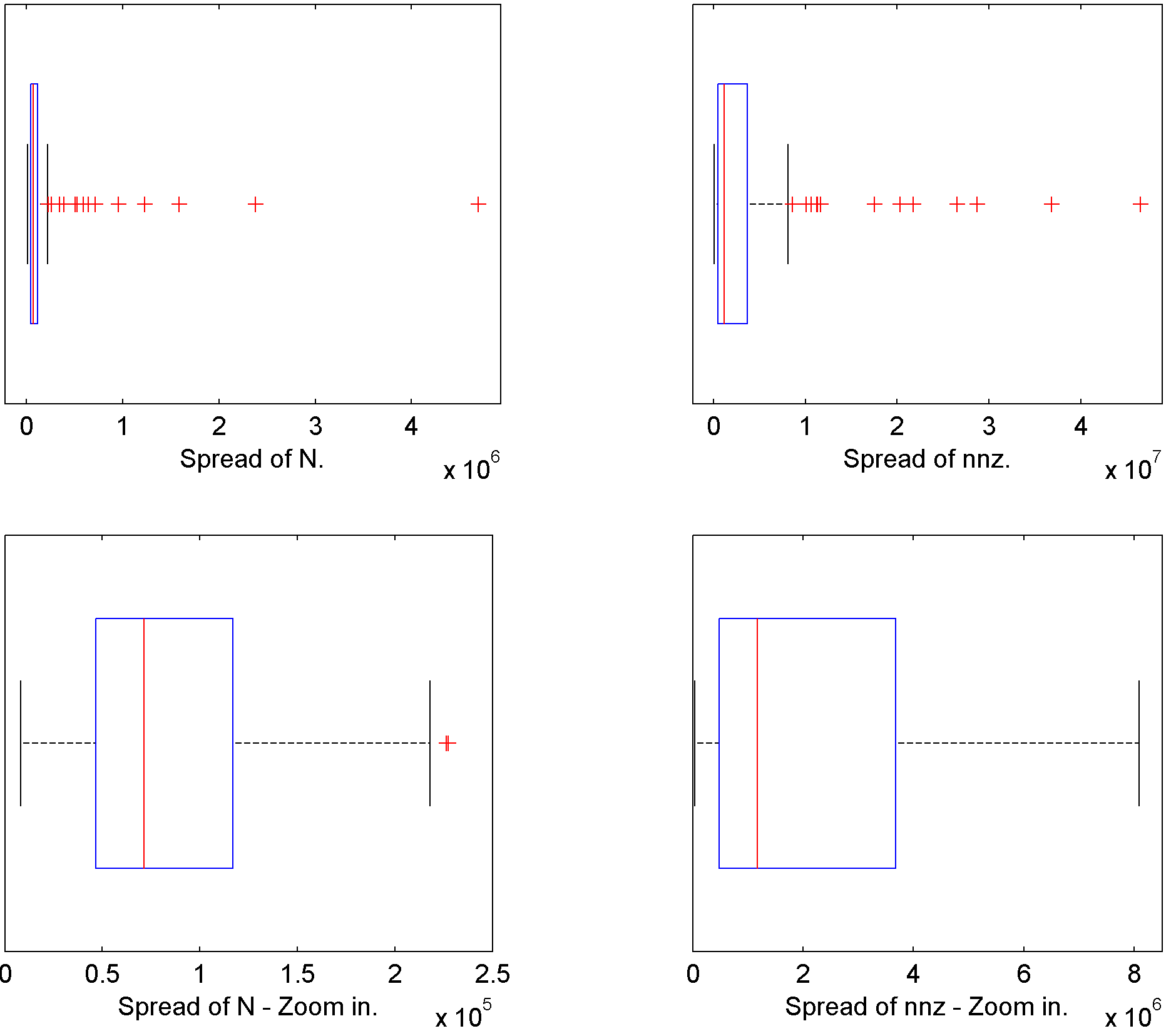}}
	\caption{Statistical information regarding the dimension $N$ and number of nonzeros $nnz$ for the 114 coefficient matrices used to compare {\SpikeHyb}, {\Pardiso}, {\SuperLU}, and {\MUMPS}.}
	\label{fig:infoBenchmarkMatrices}
\end{figure}

A set of 114 matrices, of which 105 are from the Florida matrix collection, is used herein to compare the robustness and time to solution of {\SpikeHyb}, {\Pardiso}, {\SuperLU}, and {\MUMPS}. This set of matrices was selected on the following basis: at least one of the four solvers can retrieve the solution ${\bx}$ within 1\% relative accuracy. For a sparse linear system ${\bA \bx = \bb}$, this relative accuracy was measured as follows. An exact solution ${\bf x}^\star$ was first chosen and then the right-hand side was set to $\bb = \bA {\bf x}^\star$. Each sparse linear solver attempted to produce an approximation $\bx$ of the solution ${\bf x}^\star$. If this approximation satisfied $||\bx-{\bf x}^\star||_2/||{\bf x}^\star||_2 \leq 0.01$, then the solve was considered to have been successful. Given that {\SpikeHyb} is an iterative solver, its initial guess is always $\bx^{(0)}={\bf 0}_N$. Although in many instances the initial guess can be selected to be relatively close the actual solution, this situation is avoided here by choosing ${\bf x}^\star$ far from the aforementioned initial guess. Specifically, ${\bf x}^\star$  had its entries roughly distributed on a parabola starting from 1.0 as the first entry, approaching the value $400$ at $N/2$, and decreasing to 1.0 for the $N^{th}$ and last entry of ${\bf x}^\star$.  The statistical results reported in this section draw on raw data provided in the Appendix in Table \ref{t:rawDataComp-vs-CPUsols}.
Figure~\ref{fig:infoBenchmarkMatrices} employs a median-quartile method to measure the statistical spread of the 114 matrices used in this sparse solver comparison. In terms of size, $N$ is between \SI{8192}{} and \SI{4690002}{}. In terms of nonzeros, $nnz$ is between \SI{41746}{} and \SI{46522475}{}. The median for $N$ is \SI{71328}{}. The median for $nnz$ is \SI{1167967}{}.

% -----------------
% -----------------

On the robustness side, {\SpikeHyb} failed to solve 28 linear systems. In 23 cases, {\SaP} ran out of GPU global memory. In the remaining five cases, {\SpikeHyb} failed to converge. The rest of the solvers failed as follows: {\Pardiso} 40 times, {\SuperLU} 22 times, and {\MUMPS} 35 times.  These results should be qualified as follows. The GPU card had 6 GB of GDDR5-type memory. Given that in its current implementation {\SpikeHyb} is an in-core solver, it does not swap data in and out of the GPU. Consequently, it ran 23 times against this memory-size hard constraint. This issue can be partially alleviated by considering a better GPU card. Indeed, there are cards that have as much as 24 GB of global memory, which still comes short of the 64 GB of RAM that {\Pardiso}, {\SuperLU}, and {\MUMPS} could tap into. Secondly, the {\Pardiso}, {\SuperLU}, and {\MUMPS} solvers were used with default setting. Adjusting parameters that control these solvers' solution process would likely increase their success rate.

Interestingly, for the 114 linear systems considered there was a perfect negative correlation between speed and robustness. {\Pardiso} was the fastest, followed by {\MUMPS}, then {\SaP}, and finally {\SuperLU}. Of the 57 linear systems solved both by {\SaP} and {\Pardiso}, {\SaP} was faster 20 times.  Of the 71 linear systems solved both by {\SaP} and {\SuperLU}, {\SaP} was faster 38 times. Of the 60 linear systems solved both by {\SaP} and {\MUMPS}, {\SaP} was faster 27 times. Of the 60 linear systems solved both by {\Pardiso} and {\SuperLU}, {\Pardiso} was faster 60 times. Of the 57 linear systems solved both by {\SaP} and {\MUMPS}, {\Pardiso} was faster 57 times. And finally, of the 64 linear systems solved both by {\SuperLU} and {\MUMPS}, {\SuperLU} was faster 24 times. 

\begin{figure}[ht]
	\centering {\includegraphics[width=0.8\textwidth]{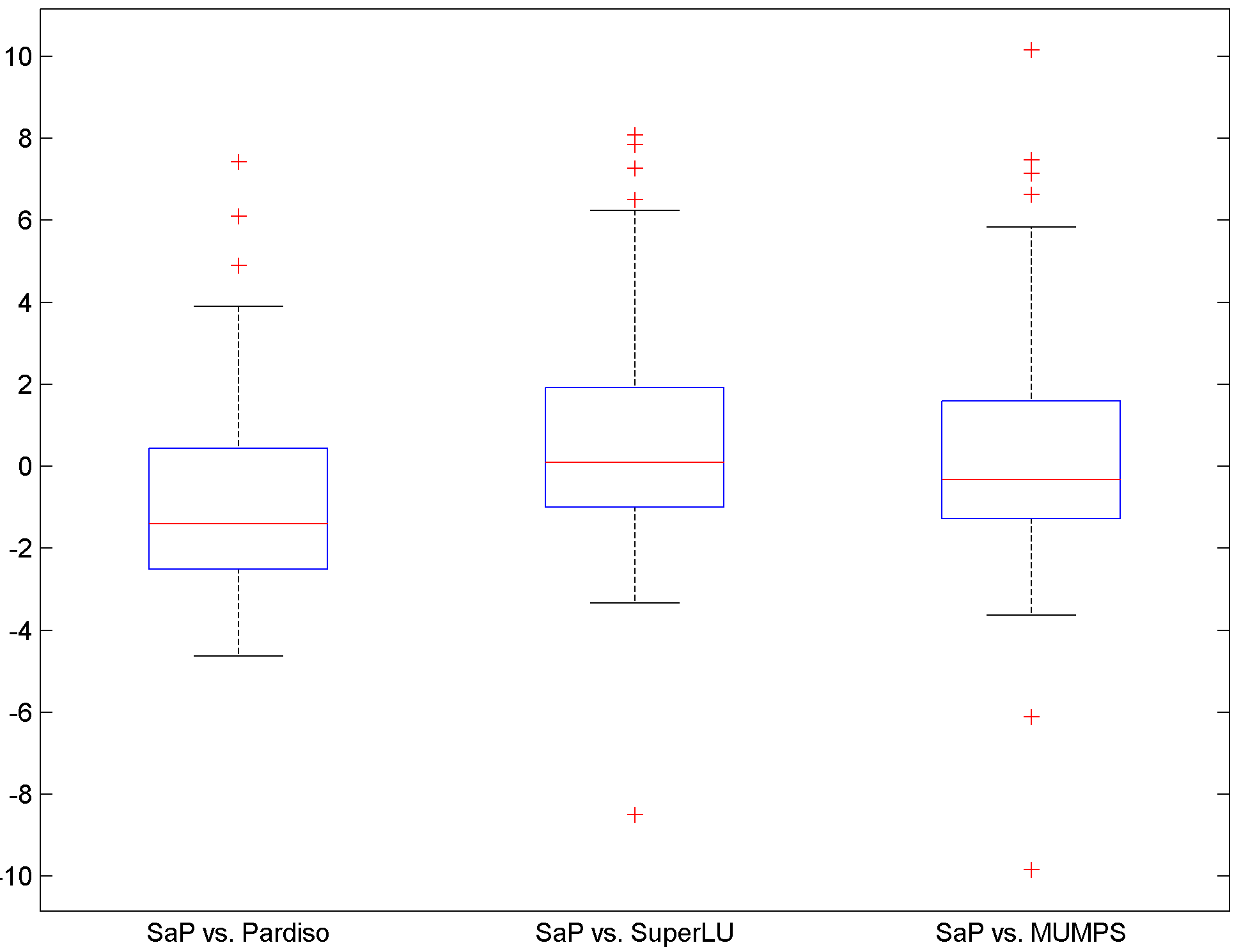}}
	\caption{Statistical spread for {\SpikeHyb}'s performance relative to that of {\Pardiso}, {\SuperLU}, and {\MUMPS}. Referring to Eq.~\ref{eq:spreadDefinition}, the results were obtained using the data sets ${\cal S}^{\SaP-\Pardiso}$ (with 57 values), ${\cal S}^{\SaP-\SuperLU}$ (71 values), and ${\cal S}^{\SaP-\MUMPS}$ (60 values).}
	\label{f:boxPlotSparseSolvers}
\end{figure}

We compare next the four solvers using a median-quartile method to measure statistical spread. Assume that $T_\alpha^{\SaP}$ and $T_\alpha^{\Pardiso}$ represent the times required by {\SpikeHyb} and {\Pardiso}, respectively, to finish test $\alpha$. A relative speedup is computed as 
\begin{equation}
{\cal S}_\alpha^{\SaP-\Pardiso} = \log_2 \frac{T_\alpha^{\Pardiso}}{T_\alpha^{\SaP}} \, ,
%\left\{ {
%	\begin{aligned}
%	\log_2 \frac{T_\alpha^{\Pardiso}}{T_\alpha^{\SaP}} &\quad \text{if} \quad  T_\alpha^{\Pardiso}>T_\alpha^{\SaP} \vspace{0.25cm}\\
%	-\log_2 \frac{T_\alpha^{\SaP}}{T_\alpha^{\Pardiso}} &\quad \text{if} \quad T_\alpha^{\SaP} \geq T_\alpha^{\Pardiso}
%	\end{aligned}
%}\right.
%\quad ,
\label{eq:spreadDefinition}
\end{equation}
\noindent with ${\cal S}_\alpha^{\SaP-\MUMPS}$ and ${\cal S}_\alpha^{\SaP-\SuperLU}$ similarly computed. These ${\cal S}_\alpha^{\SaP-\Pardiso}$ values, which can be either positive or negative, are collected in a set ${\cal S}^{\SaP-\Pardiso}$ which is used to generate a box plot in Fig.~\ref{f:boxPlotSparseSolvers}. The figure also reports results on ${\cal S}^{\SaP-\SuperLU}$, and ${\cal S}^{\SaP-\MUMPS}$. Note that the number of tests used to produce these statistical measures is different for each comparison: 57 linear systems for ${\cal S}^{\SaP-\Pardiso}$, 71 for ${\cal S}^{\SaP-\SuperLU}$, and 60 for ${\cal S}^{\SaP-\MUMPS}$. The median values for ${\cal S}^{\SaP-\Pardiso}$, ${\cal S}^{\SaP-\SuperLU}$, and ${\cal S}^{\SaP-\MUMPS}$ are $-1.4036$, $0.0934$, and $-0.3242$, respectively. These results suggest that when it finishes, {\Pardiso} can be expected to be about two times faster than {\SaP}. {\MUMPS} is marginally faster than {\SaP}, which on average can be expected to be only slightly faster than {\SuperLU}.

% -----------------
% -----------------

Red crosses are used in Fig.~\ref{f:boxPlotSparseSolvers} to show statistical outliers. Favorably, most of the {\SaP}'s outliers are large and positive. For instance, there are three linear systems for which compared to {\Pardiso}, {\SaP} finishes significantly faster, four linear systems for which it is significantly faster than {\SuperLU}, and four linear systems for which it is significantly faster than {\MUMPS}. On the flip side, there are two tests where {\SaP} runs slower than {\MUMPS} and one test where it runs significantly slower then {\SuperLU}. The results also suggest that about 50\% of the linear systems run in {\SaP} in the range between ``as fast as {\Pardiso} or two to three times slower'', 50\% of the linear systems run in {\SaP} in the range ``between four times faster to four times slower then {\SuperLU}''. Relative to {\MUMPS}, the situation is just like for {\SuperLU} if only slightly shifted towards negative territory: the second and third quartile suggest that 50\% of the linear systems run in {\SaP} in the range ``between three times faster to three times slower then {\MUMPS}''. Again, favorably for {\SaP}, the last quartile is long and reaches well into high positive values. In other words, when it beats the competition, it beats it by a large margin.

\subsubsection{Comparison against another GPU solver}
\label{sss:compareAgainst-cuSolver}
The same set of 114 matrices used in the comparison against {\Pardiso}, {\SuperLU}, and {\MUMPS} was considered to compare {\SpikeHyb} with the sparse direct QR solver in {\cuSOLVER} library~\cite{cuSOLVER}. For {\cuSOLVER}, the QR solver was run in two configurations: with or without the application of a reversed Cuthill--McKee ({\RCM}) reordering before solving the system. {\RCM} was optionally applied given that it can potentially reduce the QR factorization fill-in. {\cuSOLVER} successfully solved 45 out of 114 systems when using either configuration. There are only three linear systems: {\tt ABACUS\_shell\_ud}, {\tt ex11} and {\tt jan99jac120}, which were successfully solved by {\cuSOLVER}  but not by {\SpikeHyb}. Of the 42 systems solved both by {\SpikeHyb} and {\cuSOLVER}, {\cuSOLVER} was faster than {\SpikeHyb} in five cases. In all 69 systems {\cuSOLVER} failed to solve, the implementation ran out of memory.

\FloatBarrier

%-------------------------------------------------------------------------------
\section{Conclusions and future work}
\label{s:conclusions}
This contribution discusses parallel strategies to ($i$) solve dense banded linear systems; ($ii$) solve sparse linear systems; and ($iii$) perform matrix reorderings for diagonal boosting and bandwidth reduction. The salient feature shared by these strategies is that they are designed to run in parallel on GPU cards. BSD3 open source implementations of all these strategies are available at \cite{SaP_git,SaPWebsite} as part of a software package called {\SaP}. As far as the parallel solution of linear systems is concerned, the strategies discussed are in-core; i.e., there is no host-device, CPU-GPU, memory swapping, which somewhat limits the size of the problems that can be presently solved by {\SaP}. Over a broad range of dense matrix sizes and bandwidths, {\SaP} is likely to run two times faster than Intel's MKL. This conclusion should be modulated by hardware considerations and also the observation that the diagonal dominance of the dense banded matrix is a performance factor. On the sparse linear system side, the most surprising result was the robustness of {\SaP}. Out of a set of 114 tests, most of them using matrices from the University of Florida sparse matrix collection, {\SaP} failed only 28 times, of which 23 were ``out-of-memory'' failures owing to a 6 GB limit on the size of the GPU memory. In terms of performance, {\SaP} was compared against {\Pardiso}, {\MUMPS}, and {\SuperLU}. We noticed a perfect negative correlation between robustness and time to solution: the faster a solver, the less robust it was. In this context, {\Pardiso} was the fastest, followed by {\MUMPS}, {\SaP}, and {\SuperLU}. Surprisingly, the straight split-and-parallelize strategy, without the coupling involved in the SPIKE-type strategy, emerged as the more often solution approach adopted by {\SaP}.

The implementation of {\SaP} is somewhat peculiar in that the sparse solver builds on top of the dense banded one. The sparse--to--dense transition occurs via two reorderings: one that boosts the diagonal entires and one that reduces the matrix bandwidth. Herein, they were implemented as CPU/GPU hybrid solutions which were compared against Harwell's implementations and found to be twice as fast for the diagonal boosting reordering, and of comparable speed for the bandwidth reduction. 

Many issues remain to be investigated at this point. First, given that more than 50\% of the time to solution is spent in the iterative solver, it is worth consider the techniques analyzed in \cite{SpMV-TR2015}, which sometimes double the flop rate in sparse matrix-vector multiplication operations upon changing the matrix storage scheme; i.e., moving from CSR to ELL or hybrid. Second, an out-of-core and/or multi-GPU implementation would enable {\SaP} to handle larger problems while possibly reducing time to solution. Third, the {\CM} bandwidth reduction strategy implemented is dated; spectral and/or hyper-graph partitioning for load balancing should lead to superior splitting of the coefficient matrix. 
Finally, as it stands, with the exception of parts of the matrix reordering, {\SaP} is entirely a GPU solution. It would be worth investigating how the CPU can be involved in other phases of the implementation. Such an investigation would be well justified given the imminent tight integration of the CPU and GPU memories.

%-------------------------------------------------------------------------------
\section*{Acknowledgments}
This work was funded through National Science Foundation grant SI2-SSE 1147337 and benefited from many discussions the authors had with Matt Knepley and Ahmed Sameh.

%-------------------------------------------------------------------------------
\appendix

\section{Solver comparisons raw data}
For completeness, we provide here the raw comparison data for the tested solvers which was used in generating the figures and plots in the paper.
Table~\ref{t:rawDataMatrixInfo} gives the list of tested matrices, specifying their size $N$ and number of non-zero elements $nnz$. 
Table~\ref{t:rawDataComp-vs-CPUsols} reports the run times to solution (in $ms$) for the {\SpikeHyb}, {\Pardiso}, {\SuperLU}, and {\MUMPS} solvers.
Table~\ref{t:rawDataComp-vs-GPUsols} reports the run times to solution comparison for {\SpikeHyb} and {\cuSOLVER}, the latter without or with Cuthill-McKee ({\CM}) reordering.
\begingroup
  \footnotesize
  \begin{longtable}[c]{rcrr}
  	
	\caption{Dimension $N$ and number of non-zero elements of tested matrices.}
	\label{t:rawDataMatrixInfo}
	
	% Heading on first page
	\\ \hline
	& \textbf{Name} & \textbf{N} & \textbf{nnz} \\ \hline
	\endfirsthead
	
	% Heading on subsequent pages
	\multicolumn{4}{c}{{\tablename\ \thetable{} -- continued from previous page}} \\ \hline
	& \textbf{Name} & \textbf{N} & \textbf{nnz} \\ \hline
	\endhead
	
	% Footer on previous pages
	\hline \multicolumn{4}{r}{{Continued on next page}}
	\endfoot

	% Footer on last page
	\hline
	\endlastfoot

    % TABLE DATA
		1 &                 2cubes\_sphere  &       \SI{101492}{} &      \SI{1647264}{} \\
		2 &                2D\_54019\_highK &        \SI{54019}{} &       \SI{996414}{} \\
		3 &                      a2nnsnsl   &        \SI{80016}{} &       \SI{347222}{} \\
		4 &                      a5esindl   &        \SI{60008}{} &       \SI{255004}{} \\
		5 &               ABACUS\_shell\_ud &        \SI{23412}{} &       \SI{218484}{} \\
		6 &                     af\_5\_k101 &       \SI{503625}{} &     \SI{17550675}{} \\
		7 &                       af23560   &        \SI{23560}{} &       \SI{484256}{} \\
		8 &                     ANCF31770   &        \SI{31770}{} &       \SI{183540}{} \\
		9 &                     ANCF88950   &        \SI{88950}{} &       \SI{513900}{} \\
		10 &                       apache1   &        \SI{80800}{} &       \SI{542184}{} \\
		11 &                       apache2   &       \SI{715176}{} &      \SI{4817870}{} \\
		12 &                          appu   &        \SI{14000}{} &      \SI{1853104}{} \\
		13 &                     ASIC\_100k  &        \SI{99340}{} &       \SI{954163}{} \\
		14 &                    ASIC\_100ks  &        \SI{99190}{} &       \SI{578890}{} \\
		15 &                       av41092   &        \SI{41092}{} &      \SI{1683902}{} \\
		16 &                       bayer01   &        \SI{57735}{} &       \SI{277774}{} \\
		17 &                      bcircuit   &        \SI{68902}{} &       \SI{375558}{} \\
		18 &                      bcsstk39   &        \SI{46772}{} &      \SI{2089294}{} \\
		19 &                      blockqp1   &        \SI{60012}{} &       \SI{640033}{} \\
		20 &                        bmw3\_2  &       \SI{227362}{} &     \SI{11288630}{} \\
		21 &                      bmwcra\_1  &       \SI{148770}{} &     \SI{10644002}{} \\
		22 &                         boyd1   &        \SI{93279}{} &      \SI{1211231}{} \\
		23 &                       bratu3d   &        \SI{27792}{} &       \SI{173796}{} \\
		24 &                       bundle1   &        \SI{10581}{} &       \SI{770901}{} \\
		25 &                          c-59   &        \SI{41282}{} &       \SI{480536}{} \\
		26 &                          c-61   &        \SI{43618}{} &       \SI{310016}{} \\
		27 &                          c-62   &        \SI{41731}{} &       \SI{559343}{} \\
		28 &                          cant   &        \SI{62451}{} &      \SI{4007383}{} \\
		29 &                        case39   &        \SI{40216}{} &      \SI{1042160}{} \\
		30 &                   case39\_A\_01 &        \SI{40216}{} &      \SI{1042160}{} \\
		31 &                         c-big   &       \SI{345241}{} &      \SI{2341011}{} \\
		32 &                          cfd1   &        \SI{70656}{} &      \SI{1828364}{} \\
		33 &                          cfd2   &       \SI{123440}{} &      \SI{3087898}{} \\
		34 &                     circuit\_4  &        \SI{80209}{} &       \SI{307604}{} \\
		35 &                 ckt11752\_tr\_0 &        \SI{49702}{} &       \SI{333029}{} \\
		36 &                      cont-201   &        \SI{80595}{} &       \SI{438795}{} \\
		37 &                      cont-300   &       \SI{180895}{} &       \SI{988195}{} \\
		38 &                       copter2   &        \SI{55476}{} &       \SI{759952}{} \\
		39 &                    CurlCurl\_4  &      \SI{2380515}{} &     \SI{26515867}{} \\
		40 &                       dawson5   &        \SI{51537}{} &      \SI{1010777}{} \\
		41 &                           dc1   &       \SI{116835}{} &       \SI{766396}{} \\
		42 &                      dixmaanl   &        \SI{60000}{} &       \SI{299998}{} \\
		43 &                      Dubcova2   &        \SI{65025}{} &      \SI{1030225}{} \\
		44 &                        dw8192   &         \SI{8192}{} &        \SI{41746}{} \\
		45 &                         ecl32   &        \SI{51993}{} &       \SI{380415}{} \\
		46 &                          epb3   &        \SI{84617}{} &       \SI{463625}{} \\
		47 &                          ex11   &        \SI{16614}{} &      \SI{1096948}{} \\
		48 &                          ex19   &        \SI{12005}{} &       \SI{259879}{} \\
		49 &               FEM\_3D\_thermal1 &        \SI{17880}{} &       \SI{430740}{} \\
		50 &                      filter3D   &       \SI{106437}{} &      \SI{2707179}{} \\
		51 &                      finan512   &        \SI{74752}{} &       \SI{596992}{} \\
		52 &                    G3\_circuit  &      \SI{1585478}{} &      \SI{7660826}{} \\
		53 &                      g7jac140   &        \SI{41490}{} &       \SI{565956}{} \\
		54 &                     Ga3As3H12   &        \SI{61349}{} &      \SI{5970947}{} \\
		55 &                        GaAsH6   &        \SI{61349}{} &      \SI{3381809}{} \\
		56 &                        garon2   &        \SI{13535}{} &       \SI{390607}{} \\
		57 &                    gas\_sensor  &        \SI{66917}{} &      \SI{1703365}{} \\
		58 &                      gridgena   &        \SI{48962}{} &       \SI{512084}{} \\
		59 &                    gsm\_106857  &       \SI{589446}{} &     \SI{21758924}{} \\
		60 &                           H2O   &        \SI{67024}{} &      \SI{2216736}{} \\
		61 &                      hcircuit   &       \SI{105676}{} &       \SI{513072}{} \\
		62 &                  HTC\_336\_4438 &       \SI{226340}{} &       \SI{904522}{} \\
		63 &                  ibm\_matrix\_2 &        \SI{51448}{} &      \SI{1056610}{} \\
		64 &                      inline\_1  &       \SI{503712}{} &     \SI{36816342}{} \\
		65 &                   jan99jac120   &        \SI{41374}{} &       \SI{260202}{} \\
		66 &                         ldoor   &       \SI{952203}{} &     \SI{46522475}{} \\
		67 &                        lhr10c   &        \SI{10672}{} &       \SI{232633}{} \\
		68 &                           Lin   &       \SI{256000}{} &      \SI{1766400}{} \\
		69 &                         lung2   &       \SI{109460}{} &       \SI{492564}{} \\
		70 &                      mario002   &       \SI{389874}{} &      \SI{2101242}{} \\
		71 &                   mark3jac100   &        \SI{45769}{} &       \SI{285215}{} \\
		72 &                   mark3jac140   &        \SI{64089}{} &       \SI{399735}{} \\
		73 &                      matrix\_9  &       \SI{103430}{} &      \SI{2121550}{} \\
		74 &                      minsurfo   &        \SI{40806}{} &       \SI{203622}{} \\
		75 &                      msc23052   &        \SI{23052}{} &      \SI{1154814}{} \\
		76 &                      ncvxbqp1   &        \SI{50000}{} &       \SI{349968}{} \\
		77 &                         nd24k   &        \SI{72000}{} &     \SI{28715634}{} \\
		78 &                 NetANCF40by40   &        \SI{63603}{} &       \SI{569262}{} \\
		79 &                      offshore   &       \SI{259789}{} &      \SI{4242673}{} \\
		80 &                        oilpan   &        \SI{73752}{} &      \SI{3597188}{} \\
		81 &                      olesnik0    &        \SI{88263}{} &       \SI{744216}{} \\
		82 &                     OPF\_10000   &        \SI{43887}{} &       \SI{467711}{} \\
		83 &                 parabolic\_fem   &       \SI{525825}{} &      \SI{3674625}{} \\
		84 &                       pdb1HYS    &        \SI{36417}{} &      \SI{4344765}{} \\
		85 &                    poisson3Db    &        \SI{85623}{} &      \SI{2374949}{} \\
		86 &                          pwtk    &       \SI{217918}{} &     \SI{11634424}{} \\
		87 &                         qa8fk    &        \SI{66127}{} &      \SI{1660579}{} \\
		88 &                         qa8fm    &        \SI{66127}{} &      \SI{1660579}{} \\
		89 &                      raefsky4    &        \SI{19779}{} &      \SI{1328611}{} \\
		90 &                    rail\_79841   &        \SI{79841}{} &       \SI{553921}{} \\
		91 &                       rajat30    &      \SI{643994}{}  &     \SI{6175377}{}  \\ 
		92 &                       rajat31    &      \SI{4690002}{} &     \SI{20316253}{} \\
		93 &                         rma10    &        \SI{46835}{} &      \SI{2374001}{} \\
		94 &                      s3dkq4m2    &        \SI{90449}{} &      \SI{4820891}{} \\
		95 &                shallow\_water1   &        \SI{81920}{} &       \SI{327680}{} \\
		96 &                shallow\_water2   &        \SI{81920}{} &       \SI{327680}{} \\
		97 &                      ship\_003   &       \SI{121728}{} &      \SI{8086034}{} \\
		98 &                      shipsec1    &       \SI{140874}{} &      \SI{7813404}{} \\
		99 &                      shipsec5    &       \SI{179860}{} &     \SI{10113096}{} \\
		100 &                       Si34H36    &        \SI{97569}{} &      \SI{5156379}{} \\
		101 &                          SiO2    &       \SI{155331}{} &     \SI{11283503}{} \\
		102 &                      sparsine    &        \SI{50000}{} &      \SI{1548988}{} \\
		103 &                       stomach    &       \SI{213360}{} &      \SI{3021648}{} \\
		104 &                          t3dh    &        \SI{79171}{} &      \SI{4352105}{} \\
		105 &                        t3dh\_a   &        \SI{79171}{} &      \SI{4352105}{} \\
		106 &                      thermal1    &        \SI{82654}{} &       \SI{574458}{} \\
		107 &                      thermal2    &      \SI{1228045}{} &      \SI{8580313}{} \\
		108 &                        torso3    &       \SI{259156}{} &      \SI{4429042}{} \\
		109 &              TSOPF\_FS\_b162\_c4 &        \SI{40798}{} &      \SI{2398220}{} \\
		110 &              TSOPF\_FS\_b39\_c19 &        \SI{76216}{} &      \SI{1977600}{} \\
		111 &                       vanbody    &        \SI{47072}{} &      \SI{2336898}{} \\
		112 &                      venkat25    &        \SI{62424}{} &      \SI{1717792}{} \\
		113 &                        xenon1    &        \SI{48600}{} &      \SI{1181120}{} \\
		114 &                        xenon2    &       \SI{157464}{} &      \SI{3866688}{}

  \end{longtable}
\endgroup

\begingroup
  \footnotesize
  \begin{longtable}[c]{rccccc}
  	
	\caption{Run times to solution required by {\SpikeHyb}, {\Pardiso}, {\SuperLU}, and {\MUMPS}, reported in milliseconds. For {\Pardiso}, {\SuperLU}, and {\MUMPS}, a  ``-'' sign indicates an instance in which the solver failed to solve that particular linear system. When {\SpikeHyb} fails, OOM stands for ``out of memory'' and NC for ``no convergence''.} 
	\label{t:rawDataComp-vs-CPUsols} 
	
	% Heading on first page
	\\ \hline
	\multirow{2}{*}{} & 
	\multirow{2}{*}{\textbf{Name}} & 
	\multicolumn{4}{c}{Run times ($ms$)} \\ \cline{3-6}
	& & \textbf{\SpikeHyb} & \textbf{\Pardiso} & \textbf{\SuperLU} & \textbf{\MUMPS} \\ \hline
	\endfirsthead
	
	% Heading on subsequent pages
	\multicolumn{6}{c}{{\tablename\ \thetable{} -- continued from previous page}} \\ \hline
	\multirow{2}{*}{} & 
	\multirow{2}{*}{\textbf{Name}} & 
	\multicolumn{4}{c}{Run times ($ms$)} \\ \cline{3-6}
	& & \textbf{\SpikeHyb} & \textbf{\Pardiso} & \textbf{\SuperLU} & \textbf{\MUMPS} \\ \hline 
	\endhead
	
	% Footer on previous pages
	\hline \multicolumn{6}{r}{{Continued on next page}}
	\endfoot

	% Footer on last page
	\hline
	\endlastfoot

    % TABLE DATA
		1 &                 2cubes\_sphere  &    \SciNum{189.921} &   \SciNum{2829.898} &      \SciNum{14300} &    \SciNum{18829.7} \\
		2 &                2D\_54019\_highK &    \SciNum{3805.16} &         - &         - &         - \\
		3 &                      a2nnsnsl   &         OOM &    \SciNum{328.273} &        \SciNum{500} &         - \\
		4 &                      a5esindl   &         OOM &    \SciNum{148.044} &        \SciNum{240} &         - \\
		5 &               ABACUS\_shell\_ud &         NC &         - &        \SciNum{230} &    \SciNum{219.637} \\
		6 &                     af\_5\_k101 &    \SciNum{20587.8} &   \SciNum{3639.413} &      \SciNum{49260} &    \SciNum{16472.9} \\
		7 &                       af23560   &    \SciNum{727.269} &         - &        \SciNum{850} &    \SciNum{737.508} \\
		8 &                     ANCF31770   &    \SciNum{413.215} &    \SciNum{205.367} &        \SciNum{370} &         - \\
		9 &                     ANCF88950   &    \SciNum{1057.22} &    \SciNum{513.178} &        \SciNum{840} &         - \\
		10 &                       apache1   &    \SciNum{2642.53} &    \SciNum{676.093} &       \SciNum{2790} &    \SciNum{2107.05} \\
		11 &                       apache2   &         OOM &   \SciNum{9295.087} &     \SciNum{109090} &    \SciNum{38842.9} \\
		12 &                          appu   &    \SciNum{338.683} &  \SciNum{58160.038} &      \SciNum{91690} &         - \\
		13 &                     ASIC\_100k  &    \SciNum{688.038} &    \SciNum{628.299} &         - &    \SciNum{39202.5} \\
		14 &                    ASIC\_100ks  &    \SciNum{414.096} &    \SciNum{556.598} &         - &    \SciNum{1208.57} \\
		15 &                       av41092   &         OOM &         - &         - &    \SciNum{3756.84} \\
		16 &                       bayer01   &    \SciNum{3414.69} &         - &        \SciNum{860} &         - \\
		17 &                      bcircuit   &    \SciNum{4259.48} &    \SciNum{374.747} &       \SciNum{1250} &    \SciNum{707.761} \\
		18 &                      bcsstk39   &    \SciNum{1050.55} &    \SciNum{397.088} &       \SciNum{1370} &    \SciNum{1070.87} \\
		19 &                      blockqp1   &     \SciNum{177.75} &    \SciNum{437.189} &       \SciNum{1020} &         - \\
		20 &                        bmw3\_2  &         OOM &   \SciNum{2179.354} &         - &    \SciNum{8652.22} \\
		21 &                      bmwcra\_1  &    \SciNum{19409.1} &   \SciNum{3344.328} &      \SciNum{14210} &    \SciNum{13464.3} \\
		22 &                         boyd1   &    \SciNum{1825.01} &   \SciNum{7743.795} &      \SciNum{20960} &         - \\
		23 &                       bratu3d   &    \SciNum{301.857} &     \SciNum{394.78} &       \SciNum{1080} &         - \\
		24 &                       bundle1   &    \SciNum{180.274} &     \SciNum{98.358} &        \SciNum{170} &         - \\
		25 &                          c-59   &         OOM &    \SciNum{532.519} &       \SciNum{8970} &     \SciNum{4749.8} \\
		26 &                          c-61   &         OOM &    \SciNum{276.491} &       \SciNum{1240} &    \SciNum{844.465} \\
		27 &                          c-62   &         OOM &    \SciNum{722.833} &      \SciNum{15910} &         - \\
		28 &                          cant   &    \SciNum{1373.42} &   \SciNum{1450.747} &       \SciNum{3100} &    \SciNum{3735.14} \\
		29 &                        case39   &         OOM &         - &       \SciNum{1090} &         - \\
		30 &                   case39\_A\_01 &         OOM &         - &       \SciNum{1160} &         - \\
		31 &                         c-big   &         OOM &   \SciNum{5439.878} &         - &         - \\
		32 &                          cfd1   &    \SciNum{6849.57} &   \SciNum{1291.979} &       \SciNum{3410} &    \SciNum{3760.94} \\
		33 &                          cfd2   &    \SciNum{10378.9} &   \SciNum{2454.579} &       \SciNum{6490} &    \SciNum{9108.52} \\
		34 &                     circuit\_4  &         OOM &         - &         - &    \SciNum{331.295} \\
		35 &                 ckt11752\_tr\_0 &     \SciNum{212230} &         - &        \SciNum{590} &    \SciNum{231.066} \\
		36 &                      cont-201   &    \SciNum{1400.31} &         - &       \SciNum{1560} &         - \\
		37 &                      cont-300   &    \SciNum{7080.75} &         - &      \SciNum{25800} &         - \\
		38 &                       copter2   &    \SciNum{15834.1} &    \SciNum{744.479} &       \SciNum{4040} &    \SciNum{2815.76} \\
		39 &                    CurlCurl\_4  &         OOM &         - &       \SciNum{6920} &    \SciNum{8753.73} \\
		40 &                       dawson5   &    \SciNum{4838.91} &    \SciNum{455.242} &       \SciNum{1630} &    \SciNum{754.166} \\
		41 &                           dc1   &    \SciNum{1449.55} &         - &         - &         - \\
		42 &                      dixmaanl   &    \SciNum{399.772} &    \SciNum{173.939} &        \SciNum{490} &    \SciNum{388.066} \\
		43 &                      Dubcova2   &    \SciNum{510.195} &    \SciNum{503.546} &        \SciNum{890} &    \SciNum{741.671} \\
		44 &                        dw8192   &    \SciNum{1599.65} &         - &        \SciNum{240} &         - \\
		45 &                         ecl32   &    \SciNum{1305.52} &         - &       \SciNum{3270} &    \SciNum{4058.82} \\
		46 &                          epb3   &    \SciNum{1357.77} &         - &       \SciNum{1630} &    \SciNum{559.232} \\
		47 &                          ex11   &         NC &    \SciNum{521.179} &         - &    \SciNum{853.412} \\
		48 &                          ex19   &    \SciNum{5888.82} &         - &         - &     \SciNum{85.569} \\
		49 &               FEM\_3D\_thermal1 &    \SciNum{155.876} &    \SciNum{307.165} &        \SciNum{620} &         - \\
		50 &                      filter3D   &    \SciNum{39142.1} &   \SciNum{1581.317} &       \SciNum{4870} &    \SciNum{4343.15} \\
		51 &                      finan512   &     \SciNum{93.658} &    \SciNum{460.414} &       \SciNum{1540} &    \SciNum{585.718} \\
		52 &                    G3\_circuit  &    \SciNum{8262.62} &   \SciNum{10100.17} &   \SciNum{1.91E+06} &    \SciNum{43828.9} \\
		53 &                      g7jac140   &         OOM &         - &       \SciNum{2410} &    \SciNum{3750.34} \\
		54 &                     Ga3As3H12   &     \SciNum{378042} &  \SciNum{35275.139} &     \SciNum{183810} &     \SciNum{475071} \\
		55 &                        GaAsH6   &     \SciNum{115745} &  \SciNum{37095.899} &     \SciNum{176620} &     \SciNum{515340} \\
		56 &                        garon2   &    \SciNum{292.805} &    \SciNum{137.627} &        \SciNum{290} &    \SciNum{166.501} \\
		57 &                    gas\_sensor  &    \SciNum{4364.93} &   \SciNum{1305.603} &       \SciNum{5430} &    \SciNum{6521.87} \\
		58 &                      gridgena   &    \SciNum{1043.03} &    \SciNum{332.296} &        \SciNum{600} &    \SciNum{528.663} \\
		59 &                    gsm\_106857  &         OOM &   \SciNum{7766.276} &         - &    \SciNum{23950.8} \\
		60 &                           H2O   &     \SciNum{1092.8} &  \SciNum{32746.344} &     \SciNum{168170} &         - \\
		61 &                      hcircuit   &    \SciNum{5422.95} &         - &        \SciNum{570} &         - \\
		62 &                  HTC\_336\_4438 &         OOM &         - &       \SciNum{3970} &    \SciNum{677.915} \\
		63 &                  ibm\_matrix\_2 &      \SciNum{14775} &         - &       \SciNum{3760} &         - \\
		64 &                      inline\_1  &         OOM &   \SciNum{9868.931} &      \SciNum{73890} &    \SciNum{36260.4} \\
		65 &                   jan99jac120   &         NC &         - &       \SciNum{1300} &    \SciNum{1146.63} \\
		66 &                         ldoor   &         OOM &    \SciNum{9607.87} &     \SciNum{474590} &    \SciNum{35178.5} \\
		67 &                        lhr10c   &    \SciNum{541.604} &         - &        \SciNum{290} &    \SciNum{165.951} \\
		68 &                           Lin   &    \SciNum{81630.1} &   \SciNum{8733.291} &      \SciNum{56220} &    \SciNum{56140.5} \\
		69 &                         lung2   &    \SciNum{383.143} &         - &       \SciNum{1240} &    \SciNum{469.259} \\
		70 &                      mario002   &         OOM &   \SciNum{1931.287} &      \SciNum{93750} &         - \\
		71 &                   mark3jac100   &      \SciNum{10075} &         - &       \SciNum{1440} &    \SciNum{4154.63} \\
		72 &                   mark3jac140   &    \SciNum{13025.5} &         - &         - &    \SciNum{7056.96} \\
		73 &                      matrix\_9  &    \SciNum{889.253} &         - &      \SciNum{22220} &         - \\
		74 &                      minsurfo   &    \SciNum{121.848} &    \SciNum{172.617} &        \SciNum{660} &    \SciNum{292.002} \\
		75 &                      msc23052   &    \SciNum{2987.89} &    \SciNum{136.297} &         - &         - \\
		76 &                      ncvxbqp1   &    \SciNum{5332.41} &    \SciNum{324.804} &       \SciNum{1040} &    \SciNum{753.516} \\
		77 &                         nd24k   &    \SciNum{4576.23} &  \SciNum{62323.806} &     \SciNum{416750} &     \SciNum{815354} \\
		78 &                 NetANCF40by40   &    \SciNum{560.758} &    \SciNum{614.949} &        \SciNum{690} &    \SciNum{646.128} \\
		79 &                      offshore   &         OOM &   \SciNum{5799.706} &      \SciNum{33380} &    \SciNum{30255.4} \\
		80 &                        oilpan   &    \SciNum{3740.37} &   \SciNum{1083.565} &       \SciNum{1250} &    \SciNum{1762.63} \\
		81 &                      olesnik0    &    \SciNum{7073.93} &         - &       \SciNum{1590} &         - \\
		82 &                     OPF\_10000   &    \SciNum{4635.22} &         - &        \SciNum{460} &    \SciNum{375.411} \\
		83 &                 parabolic\_fem   &    \SciNum{11318.6} &   \SciNum{3157.938} &     \SciNum{169450} &    \SciNum{6119.75} \\
		84 &                       pdb1HYS    &    \SciNum{4347.67} &    \SciNum{921.078} &         - &     \SciNum{3353.9} \\
		85 &                    poisson3Db    &     \SciNum{1360.9} &         - &       \SciNum{8610} &    \SciNum{10094.6} \\
		86 &                          pwtk    &      \SciNum{13553} &   \SciNum{1792.319} &       \SciNum{7380} &    \SciNum{6868.76} \\
		87 &                         qa8fk    &    \SciNum{1375.42} &         - &       \SciNum{4720} &         - \\
		88 &                         qa8fm    &    \SciNum{173.168} &   \SciNum{1236.401} &       \SciNum{4670} &    \SciNum{6683.39} \\
		89 &                      raefsky4    &    \SciNum{6230.07} &    \SciNum{267.443} &         - &         - \\
		90 &                    rail\_79841   &    \SciNum{1402.15} &    \SciNum{411.543} &        \SciNum{730} &    \SciNum{685.243} \\
		91 &                       rajat30    &    \SciNum{6413.75} &         -           &         - &         - \\
		92 &                       rajat31    &    \SciNum{20217.1} &  \SciNum{31609.469} &         - &         - \\
		93 &                         rma10    &    \SciNum{1654.42} &         - &       \SciNum{1150} &    \SciNum{584.041} \\
		94 &                      s3dkq4m2    &     \SciNum{2884.1} &   \SciNum{1385.112} &       \SciNum{3710} &    \SciNum{3851.92} \\
		95 &                shallow\_water1   &     \SciNum{69.401} &    \SciNum{423.798} &       \SciNum{1320} &    \SciNum{1236.94} \\
		96 &                shallow\_water2   &     \SciNum{98.589} &    \SciNum{386.173} &       \SciNum{1300} &     \SciNum{851.78} \\
		97 &                      ship\_003   &    \SciNum{23559.4} &    \SciNum{4211.25} &      \SciNum{20840} &      \SciNum{27612} \\
		98 &                      shipsec1    &    \SciNum{49260.9} &    \SciNum{2925.65} &      \SciNum{10980} &    \SciNum{12659.1} \\
		99 &                      shipsec5    &         NC &   \SciNum{3807.329} &      \SciNum{18590} &    \SciNum{19374.1} \\
		100 &                       Si34H36    &         OOM & \SciNum{111793.324} &         - &   \SciNum{1.62E+06} \\
		101 &                          SiO2    &    \SciNum{5195.99} & \SciNum{354351.256} &         - &   \SciNum{5.94E+06} \\
		102 &                      sparsine    &         NC &  \SciNum{57603.649} &     \SciNum{245040} &     \SciNum{521780} \\
		103 &                       stomach    &     \SciNum{707.41} &         - &      \SciNum{25190} &     \SciNum{100146} \\
		104 &                          t3dh    &    \SciNum{14588.5} &         - &      \SciNum{15390} &         - \\
		105 &                        t3dh\_a   &    \SciNum{14622.2} &         - &      \SciNum{15600} &         - \\
		106 &                      thermal1    &    \SciNum{1477.23} &    \SciNum{408.663} &        \SciNum{770} &     \SciNum{873.28} \\
		107 &                      thermal2    &     \SciNum{148224} &   \SciNum{8112.395} &         - &    \SciNum{17589.4} \\
		108 &                        torso3    &    \SciNum{5410.28} &         - &         - &    \SciNum{67610.4} \\
		109 &              TSOPF\_FS\_b162\_c4 &         OOM &         - &       \SciNum{4830} &         - \\
		110 &              TSOPF\_FS\_b39\_c19 &         OOM &         - &       \SciNum{2900} &         - \\
		111 &                       vanbody    &    \SciNum{5213.21} &    \SciNum{354.307} &         - &    \SciNum{803.545} \\
		112 &                      venkat25    &    \SciNum{4182.23} &         - &       \SciNum{1160} &     \SciNum{576.78} \\
		113 &                        xenon1    &    \SciNum{4086.07} &   \SciNum{1006.297} &       \SciNum{2240} &    \SciNum{2559.61} \\
		114 &                        xenon2    &    \SciNum{3354.01} &   \SciNum{4459.362} &      \SciNum{12940} &    \SciNum{16801.1}

  \end{longtable}
\endgroup

\begingroup
  \footnotesize
  \begin{longtable}[c]{rcccc}
	
	\caption{Run times to solution required by {\SpikeHyb} and {\cuSOLVER}, reported in milliseconds. A ``-'' sign indicates a solver failure in solving a certain linear system.} 
	\label{t:rawDataComp-vs-GPUsols} 
	
	% Heading on first page
	\\ \hline
	\multirow{2}{*}{} & \multirow{2}{*}{\textbf{Name}} & \multirow{2}{*}{\textbf{\SpikeHyb}} & \multicolumn{2}{c}{\textbf{\cuSOLVER}} \\ \cline{4-5}
	& & & \textbf{w/o {\CM}} & \textbf{w/ {\CM}} \\ \hline 
	\endfirsthead
	
	% Heading on subsequent pages
	\multicolumn{5}{c}{{\tablename\ \thetable{} -- continued from previous page}} \\ \hline
	\multirow{2}{*}{} & \multirow{2}{*}{\textbf{Name}} & \multirow{2}{*}{\textbf{\SpikeHyb}} & \multicolumn{2}{c}{\textbf{\cuSOLVER}} \\ \cline{4-5}
	& & & \textbf{w/o {\CM}} & \textbf{w/ {\CM}} \\ \hline 
	\endhead
	
	% Footer on previous pages
	\hline \multicolumn{5}{r}{{Continued on next page}}
	\endfoot
	
	% Footer on last page
	\hline
	\endlastfoot
	
	% TABLE DATA
		1 &                 2cubes\_sphere   &    \SciNum{189.921} &  -            &  -  \\
		2 &                2D\_54019\_highK  &    \SciNum{3805.16} &  -            &  -  \\
		3 &                      a2nnsnsl    &         - &  -            &  -  \\
		4 &                      a5esindl    &         - &  -            &  -  \\
		5 &               ABACUS\_shell\_ud  &         - & ${\SciNum{1370.53}}$    &  -  \\
		6 &                     af\_5\_k101  &    \SciNum{20587.8} &  -            &  -  \\
		7 &                       af23560    &    \SciNum{727.269} & ${\SciNum{3397.82}}$    & ${\SciNum{3576.12}}$  \\
		8 &                     ANCF31770    &    \SciNum{413.215} & ${\SciNum{1120.17}}$    & ${\SciNum{112789}}$  \\
		9 &                     ANCF88950    &    \SciNum{1057.22} & ${\SciNum{5000.44}}$    &  -  \\
		10 &                       apache1   &    \SciNum{2642.53} & ${\SciNum{25069.5}}$    &  -  \\
		11 &                       apache2   &         - &  -            &  -  \\
		12 &                          appu   &    \SciNum{338.683} &  -            &  -  \\
		13 &                     ASIC\_100k  &    \SciNum{688.038} &  -            &  -  \\
		14 &                    ASIC\_100ks  &    \SciNum{414.096} &  -            &  -  \\
		15 &                       av41092   &         - &  -            &  -  \\
		16 &                       bayer01   &    \SciNum{3414.69} &  -            &  -  \\
		17 &                      bcircuit   &    \SciNum{4259.48} & ${\SciNum{5950.69}}$    &  -  \\
		18 &                      bcsstk39   &    \SciNum{1050.55} & ${\SciNum{6355.74}}$    & ${\SciNum{4761.05}}$ \\
		19 &                      blockqp1   &     \SciNum{177.75} &  -            &  -  \\
		20 &                        bmw3\_2  &         - &  -            &  -  \\
		21 &                      bmwcra\_1  &    \SciNum{19409.1} &  -            &  -  \\
		22 &                         boyd1   &    \SciNum{1825.01} &  -            &  -  \\
		23 &                       bratu3d   &    \SciNum{301.857} & ${\SciNum{9900.22}}$    &  -  \\
		24 &                       bundle1   &    \SciNum{180.274} &  -            &  -  \\
		25 &                          c-59   &         - &  -            &  -  \\
		26 &                          c-61   &         - &  -            &  -  \\
		27 &                          c-62   &         - &  -            &  -  \\
		28 &                          cant   &    \SciNum{1373.42} & ${\SciNum{8741.73}}$    & ${\SciNum{7895.11}}$ \\
		29 &                        case39   &         - &  -            &  -  \\
		30 &                   case39\_A\_01 &         - &  -            &  -  \\
		31 &                         c-big   &         - &  -            &  -  \\
		32 &                          cfd1   &    \SciNum{6849.57} &  -            &  -  \\
		33 &                          cfd2   &    \SciNum{10378.9} &  -            &  -  \\
		34 &                     circuit\_4  &         - &  -            &  -  \\
		35 &                 ckt11752\_tr\_0 &     \SciNum{212230} & \SciNum{212298}         & ${\SciNum{69449.7}}$ \\
		36 &                      cont-201   &    \SciNum{1400.31} & \SciNum{1384.47}        & ${\SciNum{11094.9}}$ \\
		37 &                      cont-300   &    \SciNum{7080.75} &  -            &  -  \\
		38 &                       copter2   &    \SciNum{15834.1} &  -            &  -  \\
		39 &                    CurlCurl\_4  &         - &  -            &  -  \\
		40 &                       dawson5   &    \SciNum{4838.91} &  ${\SciNum{7835.6}}$    & ${\SciNum{18277.7}}$ \\
		41 &                           dc1   &    \SciNum{1449.55} &  -                & -           \\
		42 &                      dixmaanl   &    \SciNum{399.772} &  ${\SciNum{823.5}}$         & -           \\
		43 &                      Dubcova2   &    \SciNum{510.195} &  ${\SciNum{18646.3}}$       & -           \\
		44 &                        dw8192   &    \SciNum{1599.65} &  \SciNum{2162.4}            & \SciNum{324.913}      \\
		45 &                         ecl32   &    \SciNum{1305.52} &  ${\SciNum{50897.7}}$       & -           \\
		46 &                          epb3   &    \SciNum{1357.77} &  ${\SciNum{8472.35}}$       & ${\SciNum{4013.52}}$  \\
		47 &                          ex11   &         - &  ${\SciNum{4251.72}}$       & ${\SciNum{6154.53}}$  \\
		48 &                          ex19   &    \SciNum{5888.82} &  ${\SciNum{404.809}}$       & ${\SciNum{402.123}}$  \\
		49 &               FEM\_3D\_thermal1 &    \SciNum{155.876} &   \SciNum{25960.2}          & \SciNum{2437.79}      \\
		50 &                      filter3D   &    \SciNum{39142.1} &   -               & -           \\
		51 &                      finan512   &     \SciNum{93.658} &  ${\SciNum{4192.07}}$       & -           \\
		52 &                    G3\_circuit  &    \SciNum{8262.62} &   -               & -           \\
		53 &                      g7jac140   &         - &   -               & -           \\
		54 &                     Ga3As3H12   &     \SciNum{378042} &   -               & -           \\
		55 &                        GaAsH6   &     \SciNum{115745} &   -               & -           \\
		56 &                        garon2   &    \SciNum{292.805} &  ${\SciNum{1168.58}}$       & ${\SciNum{233947}}$   \\
		57 &                    gas\_sensor  &    \SciNum{4364.93} &   -               & -           \\
		58 &                      gridgena   &    \SciNum{1043.03} &  ${\SciNum{4314.77}}$       & ${\SciNum{7840.44}}$  \\
		59 &                    gsm\_106857  &         - &   -               & -           \\
		60 &                           H2O   &     \SciNum{1092.8} &   -               & -           \\
		61 &                      hcircuit   &    \SciNum{5422.95} &  ${\SciNum{39892.1}}$       & -           \\
		62 &                  HTC\_336\_4438 &         - &   -               & -           \\
		63 &                  ibm\_matrix\_2 &      \SciNum{14775} &   -               & -           \\
		64 &                      inline\_1  &         - &   -               & -           \\
		65 &                   jan99jac120   &         - &  ${\SciNum{129512}}$        & ${\SciNum{115971}}$   \\
		66 &                         ldoor   &         - &   -               & -           \\
		67 &                        lhr10c   &    \SciNum{541.604} &  ${\SciNum{45492.6}}$       & ${\SciNum{26462.3}}$  \\
		68 &                           Lin   &    \SciNum{81630.1} &   -               & -           \\
		69 &                         lung2   &    \SciNum{383.143} &  ${\SciNum{2628.11}}$       & ${\SciNum{265772}}$   \\
		70 &                      mario002   &         - &   -               & -           \\
		71 &                   mark3jac100   &      \SciNum{10075} &  ${\SciNum{40681.1}}$       & ${\SciNum{12738.4}}$  \\
		72 &                   mark3jac140   &    \SciNum{13025.5} &  ${\SciNum{59245.1}}$       & ${\SciNum{18035.8}}$  \\
		73 &                      matrix\_9  &    \SciNum{889.253} &   -               & -           \\
		74 &                      minsurfo   &    \SciNum{121.848} &  ${\SciNum{2393.94}}$       & ${\SciNum{3478.09}}$  \\
		75 &                      msc23052   &    \SciNum{2987.89} &   -               & \SciNum{12199.6}      \\ 
		76 &                      ncvxbqp1   &    \SciNum{5332.41} & ${\SciNum{35800.8}}$        & -           \\
		77 &                         nd24k   &    \SciNum{4576.23} &   -               & -           \\
		78 &                 NetANCF40by40   &    \SciNum{560.758} & ${\SciNum{12190.9}}$        & -           \\
		79 &                      offshore   &         - &   -               & -           \\
		80 &                        oilpan   &    \SciNum{3740.37} &   -               & ${\SciNum{73219.1}}$  \\
		81 &                      olesnik0     &    \SciNum{7073.93} &    -                & -           \\
		82 &                     OPF\_10000    &    \SciNum{4635.22} &   ${\SciNum{1250.22}}$        & ${\SciNum{10745.8}}$  \\
		83 &                 parabolic\_fem    &    \SciNum{11318.6} &    -                & -           \\
		84 &                       pdb1HYS     &    \SciNum{4347.67} &   ${\SciNum{45693.3}}$        & -           \\
		85 &                    poisson3Db     &     \SciNum{1360.9} &    -                & -           \\
		86 &                          pwtk     &      \SciNum{13553} &    -                & -           \\
		87 &                         qa8fk     &    \SciNum{1375.42} &    -                & -           \\
		88 &                         qa8fm     &    \SciNum{173.168} &   -                 & ${\SciNum{50961.4}}$  \\
		89 &                      raefsky4     &    \SciNum{6230.07} &    -                & -           \\
		90 &                    rail\_79841    &    \SciNum{1402.15} &   ${\SciNum{9076.79}}$        & -           \\
		91 &                       rajat30     &         - &    -                & -           \\
		92 &                       rajat31     &    \SciNum{20217.1} &    -                & -           \\
		93 &                         rma10     &    \SciNum{1654.42} &   ${\SciNum{4818.06}}$        & -           \\
		94 &                      s3dkq4m2     &     \SciNum{2884.1} &    -                & ${\SciNum{29197.8}}$  \\
		95 &                shallow\_water1    &     \SciNum{69.401} &   ${\SciNum{11180.2}}$        & ${\SciNum{79228.3}}$  \\
		96 &                shallow\_water2    &     \SciNum{98.589} &   ${\SciNum{11006.5}}$        & ${\SciNum{78997.1}}$  \\
		97 &                      ship\_003    &    \SciNum{23559.4} &    -                & -           \\
		98 &                      shipsec1     &    \SciNum{49260.9} &    -                & -           \\
		99 &                      shipsec5     &         - &    -                & -           \\
		100 &                       Si34H36    &         - &    -                & -           \\
		101 &                          SiO2    &    \SciNum{5195.99} &    -                & -           \\
		102 &                      sparsine    &         - &    -                & -           \\
		103 &                       stomach    &     \SciNum{707.41} &    -                & -           \\
		104 &                          t3dh    &    \SciNum{14588.5} &    -                & -           \\
		105 &                        t3dh\_a   &    \SciNum{14622.2} &    -                & -           \\
		106 &                      thermal1    &    \SciNum{1477.23} &  ${\SciNum{5194.65}}$         & -           \\
		107 &                      thermal2    &     \SciNum{148224} &    -                & -           \\
		108 &                        torso3    &    \SciNum{5410.28} &    -                & -           \\
		109 &              TSOPF\_FS\_b162\_c4 &         - &    -                & -           \\
		110 &              TSOPF\_FS\_b39\_c19 &         - &    -                & -           \\
		111 &                       vanbody    &    \SciNum{5213.21} &    -                & -           \\
		112 &                      venkat25    &    \SciNum{4182.23} &  ${\SciNum{31379.5}}$         & ${\SciNum{214097}}$   \\
		113 &                        xenon1    &    \SciNum{4086.07} &    \SciNum{67085.7}           & -           \\
		114 &                        xenon2    &    \SciNum{3354.01} &    -                & -      
	
  \end{longtable}
\endgroup

\bibliographystyle{siam}
\bibliography{refMBS,refsLinAlgebra,compSciRefs,stochasticOptimization}

\end{document}